\documentclass[pra,twocolumn,showpacs,floatfix,a4paper]{revtex4}
\usepackage{graphicx,bm,amsmath}

\begin{document}
\title{Cross phase modulation in a five--level atomic medium: Semiclassical theory}
\author{Carlo Ottaviani}
\email{carlo.ottaviani@unicam.it}
\author{Stojan Rebi\'{c}}
\author{David Vitali}
\author{Paolo Tombesi}
\affiliation{Dipartimento di Fisica Universit\`a di Camerino, I-62032 Camerino, Italy}

\date{\today}
\begin{abstract}
The interaction of a five-level atomic system involving electromagnetically induced transparency with four light fields is investigated.
Two different light-atom configurations are considered, and their efficiency in generating large nonlinear cross-phase shifts compared.
The dispersive properties of those schemes are analyzed in detail, and the conditions leading to group velocity matching for two of the light
fields are identified. An analytical
treatment based on amplitude equations is used in order to obtain approximate solutions for the susceptibilities, which are shown to fit well
with the numerical solution of the full Bloch equations in a large parameter region.
\end{abstract}

\pacs{42.50.Gy, 42.65.-k, 03.67.Hk}

\maketitle

\section{Introduction}

An efficient cross-phase modulation (XPM) in quantum and semiclassical regimes is both interesting and useful in many possible applications,
such as those in optical communications \cite{Schmidt98}, optical Kerr shutters~\cite{Boyd92}, quantum non-demolition measurements~\cite{gr}
and quantum phase gates~\cite{NielsenChuang}. In all of these, but the last two especially, a large XPM is desirable for low pump powers and high sensitivities.

\begin{figure}[t]
\begin{center}
\includegraphics[width=3.0in]{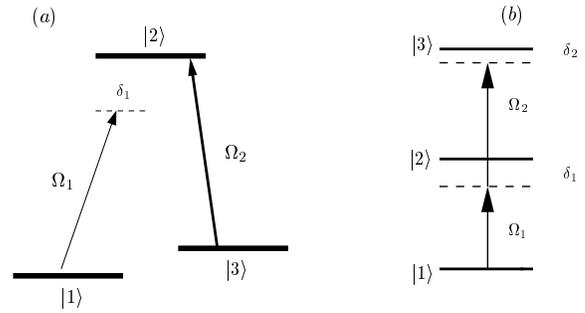}
\caption{ (a) Three-level $\Lambda$--scheme for EIT. Transitions are driven by the probe and coupling fields,
with Rabi frequency $\Omega_{1}$ and $\Omega_{2}$ respectively. When the probe detuning $\delta_1$ matches the two photon Raman-resonance condition
with the coupling field, the atomic medium becomes transparent for the probe field.
(b) Three-level cascade configuration. To obtain significant nonlinear effects for the probe field with Rabi frequency $\omega_{1}$,
a large detuning $\delta_1$ from the intermediate level $|2\rangle$ is necessary.}\label{fig1}
\end{center}
\end{figure}

In a standard three-level cascade scheme, shown in Fig.~\ref{fig1}b, nonlinear effects are obtained alongside absorption, which increase as the fields
are tuned closer to the atomic transition~\cite{ima}. To reduce the absorption to an acceptable level, light fields need to be strongly detuned
from the intermediate atomic level $|2\rangle$, simultaneously reducing however the size of the nonlinearity,
since both are inversely proportional to the square of the detuning.

Extensive studies aimed at avoiding this problem have been performed in recent years. A promising candidate emerged with the use of quantum coherence
effects in the interaction of light with multilevel atoms. Coherent population trapping (CPT)~\cite{ari} and
in particular the related effect of electromagnetically induced transparency (EIT)~\cite{eit,fleischhauerRMP} have been studied theoretically~\cite{ima,nature,harhau}
and experimentally~\cite{gr,hau,phil,zhu} in various energy-level schemes based on a generic $\Lambda$-scheme (see Fig.~\ref{fig1}a).
At resonance ($\delta_1=0$ in Fig.~\ref{fig1}a), the presence of the coupling field (with Rabi frequency $\Omega_2$)
cancels, by destructive interference, the absorption on the probe transition (with Rabi frequency $\Omega_1$),
and renders the medium transparent for the probe beam. A more general condition for EIT is
two-photon resonance, a condition that is satisfied when the frequency difference between the fields matches the energy gap between
levels $|1\rangle$ and $|3\rangle$. However, on the exact EIT resonance, probe field decouples from the atoms, making the dynamics purely linear.

Optical nonlinearities in a multilevel atomic or molecular system in the presence of EIT,
usually arise by one of the two following related mechanisms. One is to violate the strict
two-photon resonance condition, with a frequency mismatch smaller than the width of the transparency
window~\cite{gr,nostro,tripod}. Alternatively, one can add additional energy level(s) in order to induce an ac-Stark shift and effectively tune the signal
out of resonance~\cite{ima,kuri,harhau}. Both mechanisms result in large nonlinearities, accompanied by very weak absorption. Recently,
the so-called \emph{M}-scheme, shown in Fig.~\ref{fig2}, has been studied and proposed as a promising source of giant nonlinearities that can be
utilized for XPM~\cite{zub,nostro}. The double $\Lambda$ nature of this \emph{M} configuration offers the opportunity of a simultaneous group velocity
reduction for pulses propagating inside the atomic sample. Group velocity matching, originally pointed out by Lukin and Imamo\u{g}lu~\cite{luima},
is important to obtain a large XPM. In fact, it has been shown by Harris and Hau~\cite{harhau} that if equal group velocity reduction is not
achieved for both fields, the nonlinear phase accumulation will saturate at a certain constant value. The consequence is that increasing the length
of the sample in which the nonlinear interaction takes place is not useful. On the other hand, if group velocities are equal, the nonlinear
phase accumulation becomes linear in the interaction length~\cite{luima} and it may become very large.

A large cross Kerr phase shift is very useful for photonic-based implementations of quantum information (QI) processing systems \cite{NielsenChuang,prlno}.
In fact, a fundamental building block for quantum information processing is the quantum phase gate (QPG).
In a QPG, one qubit gets a phase conditional to the other qubit state according to the transformation~\cite{NielsenChuang,Lloyd} $|i\rangle _{1}|j\rangle _{2}
\rightarrow \exp\left\{i \phi_{ij} \right\}|i\rangle _{1}|j\rangle _{2} $ where $\{i,j\}=0,1$ denote the logical qubit bases.
This gate is universal when the conditional phase shift (CPS)
\begin{equation}\label{eq:def_cps}
\phi = \phi_{11} + \phi_{00} -\phi_{10} -\phi_{01},
\end{equation}
is nonzero, and it is equivalent to a CNOT gate up to local unitary transformations when $\phi=\pi$ \cite{Lloyd,NielsenChuang}.

To obtain a CPS of $\phi=\pi$, one looks for a strong interaction between qubits, ideally accompanied by weak decoherence. Photons are a particularly
attractive choice for qubits due to their robustness against decoherence during the processing and transmission of information. This feature should ideally
permit the transmission of the quantum information stored in very weak quantum pulses over very long distances with a negligibly small reduction of the initial signal.
There is however an important difficulty in the implementation of an all--optical--QPG: to process the information one needs strong
photon--photon interaction. In fact, to implement QI with photons, a nonlinear interaction is needed either to build a two-photon gate
operation~\cite{Turchette95,nostro,tripod,noipra} or at the detection stage in linear optics quantum computation~\cite{knill}.
It should also be mentioned that the generation of single-photons (which is also necessary in linear optics quantum computation) also relies on nonlinear interactions.

In this paper we perform a semiclassical analysis of the interaction of light with atoms in the \emph{M} configuration, in which the amplitude
of the four fields involved is described in terms of the corresponding Rabi frequency.
The aim is to estimate the effects of noise sources, such as dephasing and spontaneous emission, both on the nonlinear interaction and on group velocity
matching. The semiclassical regime offers a clear picture of the physical aspects involved in EIT-based nonlinear optics, and well describes
a number of recent experiments~\cite{zhu,braje}. To this end we consider two different configurations of atom-field interactions,
which we will call the \emph{asymmetric} (see Fig.~\ref{fig2}) and the \emph{symmetric} (see Fig.~\ref{fig:SMS}) \emph{M} scheme.
The paper is thus composed of two main parts. In Sec.~\ref{sec:assym} we describe the physics of the asymmetric \emph{M}-scheme.
We start by defining the system and calculating the susceptibilities using an approximate treatment employing amplitude equations.
These analytical calculations are then compared with the results of the numerical solution of the full system of Bloch equations.
Finally, the conditions for group velocity matching are analyzed. In Sec.~\ref{sec:symmM} the physics of the symmetric
\emph{M}-scheme is described by following the same order as in Sec.~\ref{sec:assym}. Conclusions are drawn in Sec.~\ref{sec:conc}.

\section{The Asymmetric M Scheme}\label{sec:assym}

\subsection{The System}

\begin{figure}[t]
\begin{center}
\includegraphics[scale=0.8]{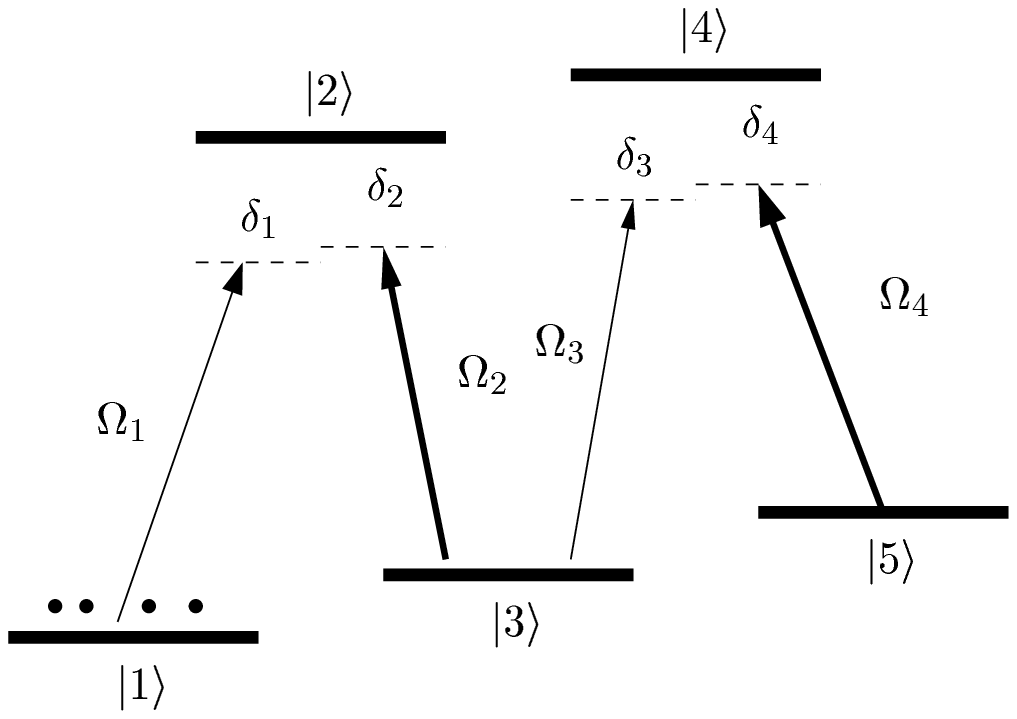}
\caption{Asymmetric \emph{M} scheme. The probe and the trigger fields, with Rabi frequencies $\Omega_{1}$ and $\Omega_{3}$ respectively,
together with the stronger pump fields, the coupler and the tuner (with Rabi frequencies $\Omega_{2}$ and $\Omega_{4}$, respectively)
drive the corresponding transitions.
All the atoms are assumed to be in state $|1\rangle $ and the detunings are defined in Eqs.~(\protect\ref{eq:det}).}\label{fig2}
\end{center}
\end{figure}

The \emph{M}-system under consideration has a double adjacent $\Lambda$ structure as shown in
Fig.~\ref{fig2}, where atoms with five levels (three ground states $|1\rangle$, $|3\rangle$, $|5\rangle$, and two excited states $|2\rangle$, $|4\rangle$)
interact with four electromagnetic fields. This configuration can be realized in Zeeman-splitted alkali atoms, such as $^{87}$Rb atoms.
The Rabi frequencies associated with the lasers driving the atomic transitions are defined as
\begin{equation}\label{eq:rabi}
\Omega_{k}=-\frac{\mu_{ij}{\mathcal E}_{k}}{\hbar},
\end{equation}\\
where ${\mathcal E}_{k}$ is the electric field amplitude, $\mu_{ij}$ is the relative dipole matrix elements induced on the
transition $|i\rangle\leftrightarrow |j\rangle$. On transitions $|3\rangle\leftrightarrow |2\rangle$ and $|5\rangle\leftrightarrow |4\rangle$
we apply two strong fields, the \emph{coupler} $\Omega_{2}$ and the \emph{tuner} $\Omega_{4}$ respectively. On the transition $|1\rangle\leftrightarrow |2\rangle$
a \emph{probe} field is applied (with $\Omega_{1}$), while on the transition $|3\rangle\leftrightarrow |4\rangle$ a \emph{trigger} field (with $\Omega_{3}$)
is applied.
In this paper, we will analyze the XPM and the group velocity matching between the \emph{probe} and the \emph{trigger} fields.
We call the scheme of Fig.~2 the \emph{asymmetric M} scheme due to the asymmetric distribution of the initial atomic population.
All the atoms are in fact assumed to be initially in state $|1\rangle$ so that they directly feel the effect of the probe field only, while the
effect of the trigger field is only indirect.
Due to this inherent asymmetry, the dynamics experienced by probe and trigger fields are always different, even when the corresponding parameters
(Rabi frequencies, decay rates, detunings) are equal. The symmetric version of this scheme
will be analyzed in Sec.~\ref{sec:symmM}.

The detunings $\delta_i$ (see Fig.~\ref{fig2}) are defined as follows
\begin{subequations}
\label{eq:det}
\begin{eqnarray}
E_{2}-E_{1}&=&\hbar\omega_{1}+\hbar\delta_{1}\\
E_{2}-E_{3}&=&\hbar\omega_{2}+\hbar\delta_{2} \\
E_{4}-E_{3}&=&\hbar\omega_{3}+\hbar\delta_{3} \\
E_{4}-E_{5}&=&\hbar\omega_{4}+\hbar\delta_{4},
\end{eqnarray}
\end{subequations}
where $E_{i}$, ($i=1,\dots,5$) is the energy of level $|i\rangle $, and $\omega_{i}$ is the frequency of the field with Rabi frequency
$\Omega_i$.

The Hamiltonian of the system is
\begin{eqnarray}\label{eq:ham1}
H_A &=& \sum_{i}^{5}E_{i}|i\rangle\langle
i|+\hbar\left(\Omega_{1}e^{-i\omega_{1}t}|2\rangle\langle1|+\Omega_{2}e^{-i\omega_{2}t}|2\rangle\langle3|\right. \nonumber\\
&&+\left. \Omega_{3}e^{-i\omega_{3}t}|4\rangle\langle3|+\Omega_{4}e^{-i\omega_{4}t}|4\rangle\langle5|+h.c.\right),
\end{eqnarray}\\
where $h.c.$ denotes the hermitian conjugate. Moving to the interaction picture with respect to the following
free Hamiltonian
\begin{eqnarray}\label{eq:hamfree}
H_{0} &=&
E_{1}|1\rangle\langle1|+(E_{2}-\hbar\delta_{1})|2\rangle\langle2|+(E_{3}-\hbar\delta_{12})|3\rangle\langle3| \nonumber \\
&& +(E_{4}-\hbar\delta_{13})|4\rangle\langle4|+(E_{5}-\hbar\delta_{14})|5\rangle\langle5| ,
\end{eqnarray}\\
where
\begin{subequations}
\label{eq:delta}
\begin{eqnarray}
\delta_{12} &=& \delta_{1}-\delta_{2},\\
\delta_{13} &=& \delta_{1}-\delta_{2}+\delta_{3},\\
\delta_{14} &=& \delta_{1}-\delta_{2}+\delta_{3}-\delta_{4},
\end{eqnarray}
\end{subequations}
we get the following effective Hamiltonian
\begin{eqnarray}\label{eq:hameff}
&& H_{eff}^{AS} =\hbar\delta_{1}|2\rangle\langle2|+\hbar\delta_{12}|3\rangle\langle3|+\hbar\delta_{13}|4\rangle\langle4|+\hbar\delta_{14}|5\rangle\langle5| \nonumber\\
&&+\hbar\Omega_{1}|2\rangle\langle1|+\hbar\Omega_{2}|2\rangle\langle3|+\hbar\Omega_{3}|4\rangle\langle3|+\hbar\Omega_{4}|4\rangle\langle5| \nonumber \\
&&+\hbar\Omega_{1}^{\star}|1\rangle\langle 2|+\hbar\Omega_{2}^{\star}|3\rangle\langle2|+\hbar\Omega_{3}^{\star}|3\rangle\langle4|+\hbar\Omega_{4}^{\star}|5\rangle\langle4|.
\end{eqnarray}

\subsection{Amplitude variables approach} \label{sec:AVasym}

We now study the dynamics driven by Eq.~(\ref{eq:hameff}). However, we have to include the effects of spontaneous emission
and dephasing, and we first treat them in a phenomenological manner by including decay rates $\Gamma_{i}^{AV}$ for each atomic
level $|i\rangle $ in the equations for the amplitude variables (AV) of the atomic wave-function. From an intuitive point of view,
for the excited levels $|2\rangle $ and $|4\rangle $ these rates describe the total spontaneous decay rates, while
for the ground states the associated decay rates describe dephasing processes \cite{fleischhauerRMP}.
Therefore, the evolution equations for the amplitudes $b_i(t)$ of the atomic state
\begin{equation}\label{eq:state}
|\psi(t)\rangle=\sum_{i=1}^{5}b_{i}(t)|i\rangle
\end{equation}
become
\begin{subequations}
\label{eq:amplisys}
\begin{eqnarray}
\dot{b}_{1}&=&-\frac{\Gamma_{1}^{AV}}{2}b_{1}-i\Omega_{1}^{\star}b_{2}, \\
\dot{b}_{2}&=&-\left(\frac{\Gamma_{2}^{AV}}{2}+i\delta_{1}\right)b_{2}-i\Omega_{1}b_{1}-i\Omega_{2}b_{3}, \\
\dot{b}_{3}&=&-\left(\frac{\Gamma_{3}^{AV}}{2}+i\delta_{12}\right)b_{3}-i\Omega^{\star}_{2}b_{2}-i\Omega^{\star}_{3}b_{4},\\
\dot{b}_{4}&=&-\left(\frac{\Gamma_{4}^{AV}}{2}+i\delta_{13}\right)b_{4}-i\Omega_{3}b_{3}-i\Omega_{4}b_{5}, \\
\dot{b}_{5}&=&-\left(\frac{\Gamma_{5}^{AV}}{2}+i\delta_{14}\right)b_{5}-i\Omega_{4}^{\star}b_{4}.
\end{eqnarray}
\end{subequations}
The system's initial state is assumed to be the ground state
$|1\rangle$. Since an efficient XPM requires a dispersive interaction, we tailor the dynamics in such a way that this initial condition
on the populations remains
essentially unaltered, even when the system reaches the steady-state, i.e.,
\begin{equation}\label{eq:adiabass}
b_{1}^{ss}\simeq 1.
\end{equation}
To this end we assume that the control field $\Omega_{2}$ is stronger then the probe field
$\Omega_{1}$, with the system being approximately on Raman resonance for the first and the second $\Lambda$ subsystems ($\delta_{1}\sim\delta_{2}$
and $\delta_{3}\sim\delta_{4}$).
Equations~(\ref{eq:amplisys}) are then solved in the steady-state. In order to get a consistent expression for the nonlinear susceptibilities one has to
consider higher order contributions to Eq.~(\ref{eq:adiabass}), which is obtained by imposing the normalization of the atomic wave-function
of Eq.~(\ref{eq:state}) at second order in $|\Omega_1/\Omega_2|$. One gets the following expression for the steady state amplitudes
\begin{subequations}
\label{eq:amplisol}
\begin{eqnarray}
b_{1}^{ss}& =& 1-\frac{\left|\Omega_{1}\right|^2 \left[\left|d_{3}\right|^2+\left|\Omega_{2}\right|^2\right]}{2\left|d_{2}d_{3}-|\Omega_{2}|^2\right|^2},\\
b_{2}^{ss}&=&\Omega_{1}\frac{d_3\left[|\Omega_{4}|^{2}-d_4 d_5\right]+|\Omega_{3}|^{2}d_5}{D_{a}}b_{1}^{ss}\\
b_{3}^{ss}&=&-\Omega_{1}\Omega^{\star}_{2}\frac{|\Omega_{4}|^{2}-d_4 d_5}{D_{a}}b_{1}^{ss}\\
b_{4}^{ss}&=&-\frac{\Omega_{1}\Omega^{\star}_{2}\Omega_{3}d_5}{D_{a}}b_{1}^{ss}\\
b_{5}^{ss}&=&\frac{\Omega_{1}\Omega^{\star}_{2}\Omega_{3}\Omega^{\star}_{4}}{D_{a}}b_{1}^{ss},
\end{eqnarray}
\end{subequations}
where we have defined
\begin{subequations}
\label{gamma}
\begin{eqnarray}
&& d_2=\delta_{1} - \imath \Gamma_{2}^{AV}/2 ,\\
&& d_3=\delta_{12} - \imath \Gamma_{3}^{AV}/2,\\
&& d_4=\delta_{13} - \imath \Gamma_{4}^{AV}/2,\\
&& d_5=\delta_{14} - \imath \Gamma_{5}^{AV}/2 ,
\end{eqnarray}
\end{subequations}
\begin{equation}\label{eq:denom}
D_{a}=\left[d_2d_3-|\Omega_{2}|^{2}\right]\left[d_4 d_5-|\Omega_{4}|^{2}\right]
-d_2d_5|\Omega_{3}|^{2}.
\end{equation}
These results can be used to determine the probe and trigger susceptibilities, which are defined as
\begin{subequations}
\label{eq:avchi}
\begin{eqnarray}
\chi_{P} &=& \frac{N \mu_{12}}{V\varepsilon_{0}{\mathcal E}_1}b_{2}^{ss}b_{1}^{ss,\star}=-\frac{N |\mu_{12}|^2}{V\hbar \varepsilon_{0}\Omega_1}
b_{2}^{ss}b_{1}^{ss,\star}, \\
\chi_{T} &=& \frac{N \mu_{34}}{V\varepsilon_{0}{\mathcal E}_3}b_{4}^{ss}b_{3}^{ss,\star}=-\frac{N |\mu_{34}|^2}{V\hbar \varepsilon_{0}\Omega_3}
b_{4}^{ss}b_{3}^{ss,\star},
\end{eqnarray}
\end{subequations}
where $N$ is the number of atoms interacting with the electromagnetic field, $V$ is the volume occupied by the gas, and
$\varepsilon_{0}$ is the vacuum dielectric constant.
Doppler broadening is neglected here. It is well known that first order Doppler effect can be cancelled by using co-propagating
laser fields \cite{ari}. In particular we emphasize that this is valid for cold atomic media in a magneto-optical trap as well as for
a standard gas cell.

Inserting Eqs.~(\ref{eq:amplisol}) into Eqs.~(\ref{eq:avchi}) and expanding
in series at the lowest orders in the probe and trigger electric fields, ${\mathcal E}_{1}$ and ${\mathcal E}_{3}$
respectively, one gets
\begin{subequations}
\label{eq:suscgen}
\begin{eqnarray}
\chi_{P}&\simeq&\chi^{(1)}_{P}+\chi^{(3,sk)}_{P}|{\mathcal E}_{1}|^{2}+\chi^{(3,ck)}_{P}|{\mathcal E}_{3}|^{2}\\
\chi_{T}&\simeq& \chi^{(3,ck)}_{T}|{\mathcal E}_{1}|^{2},
\end{eqnarray}
\end{subequations}
where we have introduced the linear susceptibility $\chi^{(1)}_{P}$, the third-order self-Kerr susceptibility $\chi^{(3,sk)}_{P}$ and the
third-order cross-Kerr susceptibilities $\chi^{(3,ck)}_{P,T}$.
Eqs.~(\ref{eq:suscgen}) clearly show the asymmetry of the scheme between the probe and trigger fields, with the latter possessing a
nonzero cross-Kerr susceptibility only. This is a consequence of the asymmetry of the population distribution,
which essentially remains in the ground state $|1\rangle$ all the time. This means that the trigger field drives a virtually
empty transition, hence the contribution to the susceptibility comes only from higher order (see~\cite{tripod} for discussion on the
link between the population distribution and a linear contribution to susceptibility). It will be shown in Sec.~\ref{sec:symmM}
that the symmetric \emph{M}-scheme brings about both a linear and a self-Kerr contribution to the trigger susceptibility.

By using Eqs.~(\ref{eq:amplisol}) and the definitions of Eqs.~(\ref{gamma}) into Eqs.~(\ref{eq:avchi}), and comparing
with Eqs.~(\ref{eq:suscgen}) at the corresponding order in the electric fields, one gets the explicit
dependence of the linear and nonlinear susceptibilities as a function of the system parameters, i.e.,
\begin{equation}
\label{eq:probelinasym}
\chi^{(1)}_{P}=\frac{N |\mu_{12}|^2}{V\hbar \varepsilon_{0}}\frac{\delta_{12}-i\Gamma_3^{AV}/2}{\left(\delta_{1}-i\Gamma_2^{AV}/2\right)
\left(\delta_{12}-i\Gamma_3^{AV}/2\right)-|\Omega_{2}|^{2}}
\end{equation}
for the probe linear susceptibility, and
\begin{widetext}
\begin{subequations}
\label{eq:suscgennonlin}
\begin{eqnarray}
&&\chi^{(3,sk)}_{P}=\frac{N |\mu_{12}|^4}{V\hbar^3 \varepsilon_{0}}
\frac{-\left(\delta_{12}-i\Gamma_3^{AV}/2\right)\left[\left|\delta_{12}-i\Gamma_3^{AV}/2\right|^2+|\Omega_{2}|^{2}\right]}{\left[\left(\delta_{1}-i\Gamma_2^{AV}/2\right)
\left(\delta_{12}-i\Gamma_3^{AV}/2\right)-|\Omega_{2}|^{2}\right]\left|\left(\delta_{1}-i\Gamma_2^{AV}/2\right)
\left(\delta_{12}-i\Gamma_3^{AV}/2\right)-|\Omega_{2}|^{2}\right|^2}, \\
&&\chi^{(3,ck)}_{P}=\frac{N |\mu_{12}|^2|\mu_{34}|^2}{V\hbar^3 \varepsilon_{0}}
\frac{|\Omega_{2}|^2\left(\delta_{14}-i\Gamma_5^{AV}/2\right)}{\left[\left(\delta_{1}-i\Gamma_2^{AV}/2\right)
\left(\delta_{12}-i\Gamma_3^{AV}/2\right)-|\Omega_{2}|^{2}\right]^2\left[\left(\delta_{13}-i\Gamma_4^{AV}/2\right)
\left(\delta_{14}-i\Gamma_5^{AV}/2\right)-\left|\Omega_4\right|^2\right]}, \\
&&\chi^{(3,ck)}_{T}=\frac{N |\mu_{12}|^2|\mu_{34}|^2}{V\hbar^3 \varepsilon_{0}}
\frac{|\Omega_{2}|^2\left(\delta_{14}-i\Gamma_5^{AV}/2\right)}{\left|\left(\delta_{1}-i\Gamma_2^{AV}/2\right)
\left(\delta_{12}-i\Gamma_3^{AV}/2\right)-|\Omega_{2}|^{2}\right|^2\left[\left(\delta_{13}-i\Gamma_4^{AV}/2\right)
\left(\delta_{14}-i\Gamma_5^{AV}/2\right)-\left|\Omega_4\right|^2\right]},
\end{eqnarray}
\end{subequations}
\end{widetext}
for the third-order nonlinear susceptibilities. The two cross-Kerr susceptibilities are identical whenever the quantity
$ \left(\delta_{1}-i\Gamma_2^{AV}/2\right) \left(\delta_{12}-i\Gamma_3^{AV}/2\right)-|\Omega_{2}|^{2}$ is (at least approximately) real.
This happens in the typical EIT situation we are considering in which $|\Omega_2|$ is large enough. In fact, when $|\Omega_{2}|^{2} \gg
\left|\left(\delta_{1}-i\Gamma_2^{AV}/2\right) \left(\delta_{12}-i\Gamma_3^{AV}/2\right)\right|$, one has \cite{nostro}
\begin{widetext}
\begin{equation}
\chi^{(3,ck)}_{P}=\chi^{(3,ck)}_{T}=\frac{N |\mu_{12}|^2|\mu_{34}|^2}{V\hbar^3 \varepsilon_{0}}
\frac{\delta_{14}-i\Gamma_5^{AV}/2}{|\Omega_{2}|^2\left[\left(\delta_{13}-i\Gamma_4^{AV}/2\right)
\left(\delta_{14}-i\Gamma_5^{AV}/2\right)-\left|\Omega_4\right|^2\right]}.
\label{eq:ktrigger2}
\end{equation}
\end{widetext}
We shall see in the next subsection that these approximate expressions for the nonlinear susceptibilities fit very well with the numerical solution
of the exact dynamics of the system.

The asymmetric \emph{M}-scheme can be seen as an extension of the four level \emph{N}-scheme introduced in Ref.~\cite{ima}, with the
addition of the coupling to an additional level $|5\rangle$
provided by the tuner field with Rabi frequency $\Omega_{4}$. In fact, it easy to check that upon setting $\Omega_{4}=0$
in Eq.~(\ref{eq:suscgennonlin}b), one recovers the third-order nonlinear susceptibility of the
four-level \emph{N}-scheme derived in Refs.~\cite{ima,zhu}).
As we will see below, the role of the tuner field is to enable a fine tuning of the group velocities,
in order to achieve group velocity matching between probe and trigger~\cite{luima,kuri,nostro}.

\subsection{Comparison with the Optical Bloch Equations} \label{sec:OBEasymMain}

We now study the dynamics of the asymmetric \emph{M} scheme of Fig.~\ref{fig2} by means of the optical Bloch
equations (OBE), which allow to describe spontaneous emission and dephasing rigorously and no more phenomenologically as in the AV treatment presented in the preceding
subsection. We consider six spontaneous decay channels, i.e., the decay
of the excited state $|2\rangle $ onto the three ground state sublevels $|1\rangle$, $|3\rangle$, and $|5\rangle$ with rates $\Gamma_{21}$, $\Gamma_{23}$
and $\Gamma_{25}$ respectively, and the corresponding decay of the excited state $|4\rangle $ onto the three sublevels
$|1\rangle$, $|3\rangle$, and $|5\rangle$ with rates $\Gamma_{41}$, $\Gamma_{43}$
and $\Gamma_{45}$ respectively. Moreover we consider dephasing of the each level $|i\rangle $ with dephasing rate $\gamma_{ii}$, so that
the master equation for the atomic density operator $\rho$ is given by
\begin{widetext}
\begin{equation}\label{eq:blochgen}
\dot{\rho}=-\frac{i}{\hbar}\left[H_{eff}^{AS},\rho\right]+\sum_{l=2,4}\sum_{k=1,3,5}\frac{\Gamma_{lk}}{2}
\left( 2\hat{\sigma}_{kl}\rho\hat{\sigma}_{kl}^{\dagger} - \hat{\sigma}_{kl}^{\dagger}\hat{\sigma}_{kl}\rho
- \rho\hat{\sigma}_{kl}^{\dagger}\hat{\sigma}_{kl} \right)+\sum_{k=1}^5\frac{\gamma_{kk}}{2}
\left( 2\hat{\sigma}_{kk}\rho\hat{\sigma}_{kk} - \hat{\sigma}_{kk}\rho
- \rho\hat{\sigma}_{kk} \right),
\end{equation}
\end{widetext}
where $H_{eff}^{AS}$ is given by Eq.~(\ref{eq:hameff}) and $\hat{\sigma}_{kl}=|k\rangle \langle l |$.
The corresponding system of OBE's for the mean values $\sigma_{ij}(t)\equiv \langle \hat{\sigma}_{ij}(t)\rangle \equiv
\rho_{ji}(t)$ is displayed in Appendix~\ref{app:OBEasym} as Eqs.~(\ref{eq:blochpopAsym}) and (\ref{eq:blochcohAsym}), where we have defined for convenience
the total decay rates
\begin{eqnarray}\label{eq:decay1}
\Gamma_{2}&=&\Gamma_{21}+\Gamma_{23}+\Gamma_{25}, \\
\Gamma_{4}&=&\Gamma_{41}+\Gamma_{43}+\Gamma_{45}, \label{eq:decay2}
\end{eqnarray}
and the composite dephasing rates
\begin{equation}\label{eq:deph}
\gamma_{ij}=\gamma_{ii}+\gamma_{jj} , \;\;\;\;i=1 \ldots 5.
\end{equation}
The OBE for the \emph{M} scheme are quite involved and less suited for an approximate analytical treatment with respect to the AV equations
of the preceding subsection. In fact, if we consider again the condition $|\Omega_1/\Omega_2| \ll 1$
and, consistently with Eq.~(\ref{eq:adiabass}), we assume that
\begin{subequations}
\label{eq:approx}
\begin{eqnarray}
\sigma_{11} &\approx& 1, \\
\sigma_{jj} &\approx& 0, \;\;\;\;  j=2,\ldots,5,
\end{eqnarray}
\end{subequations}
at the steady state, it is possible to see that by inserting Eqs.~(\ref{eq:approx}) into Eqs.~(\ref{eq:blochcohAsym}) for the
coherences, one gets a satisfactory expression for the probe linear susceptibility only. To be more specific, only the approximate linear susceptibility
fits well with the numerical solution of the OBE, while it turns out to be extremely difficult to derive analytical expressions from
Eqs.~(\ref{eq:blochpopAsym}) and (\ref{eq:blochcohAsym}) for the nonlinear susceptibilities, as simple as those
of Eqs.~(\ref{eq:suscgennonlin}), and which reproduce in the same way the exact numerical solution in the EIT regime we are studying.
Obviously, one can exactly solve analytically the OBE, but the resulting expressions are very cumbersome and not physically transparent such as those
of Eqs.~(\ref{eq:suscgennonlin}). For this reason we will analytically derive from the OBE the probe linear susceptibility only, and we will
then use the OBE only for the numerical determination of the atomic steady state. Additionally, deriving this result will enable us to draw a formal
analogy between the AV and OBE treatments (see Eqs.~(\ref{eq:reinter}) below).

The probe susceptibility is defined in terms of the atomic coherence $\sigma_{12}$ as (see also Eq.~(\ref{eq:avchi}a))
\begin{equation}\label{eq:avchi2}
\chi_{P} = \frac{N \mu_{12}}{V\varepsilon_{0}{\mathcal E}_1}\sigma_{12}=-\frac{N |\mu_{12}|^2}{V\hbar \varepsilon_{0}\Omega_1}
\sigma_{12}.
\end{equation}
Using Eqs.~(\ref{eq:approx}) and performing a series expansion at the lowest order in the probe and trigger fields,
we arrive at an approximate solution for $\sigma_{12}$,
which, inserted into Eq.~(\ref{eq:avchi2}), gives the following expression for the probe linear susceptibility
\begin{equation}\label{eq:suscprobe}
\chi^{(1)}_{P}=\frac{N |\mu_{12}|^2}{V\hbar \varepsilon_{0}}\frac{\delta_{12}-i\gamma_{13}/2}{\left[\delta_{12}-i\gamma_{13}/2\right]
\left[\delta_{1}-i\left(\Gamma_{2}+\gamma_{12}\right)/2\right]-|\Omega_{2}|^{2}}.
\end{equation}
By comparing Eq.~(\ref{eq:suscprobe}) with Eq.~(\ref{eq:probelinasym}), one can immediately see that
the AV and OBE predictions for the probe linear susceptibility coincide provided that
the phenomenological decay rates $\Gamma_i^{AV}$ are appropriately interpreted, i.e.,
\begin{subequations}
\label{eq:reinter}
\begin{eqnarray}
\Gamma_2^{AV}& \leftrightarrow &\Gamma_{2}+\gamma_{12}, \\
\Gamma_3^{AV}& \leftrightarrow &\gamma_{13}.
\end{eqnarray}
\end{subequations}
This comparison shows therefore that the AV approach provides a treatment of the atomic dynamics simpler than the OBE's approach, but roughly equivalent,
and that the intuitive interpretation of its phenomenological decay rates $\Gamma_i^{AV}$
as spontaneous emission total decay rates for the
excited states, and as dephasing rates in the case of ground state sublevels, is essentially correct, especially in the typical case
in which dephasing rates are much smaller than spontaneous emission decay rates (see Eqs.~(\ref{eq:reinter})).

\begin{figure}[t]
\begin{center}
\includegraphics[scale=0.45]{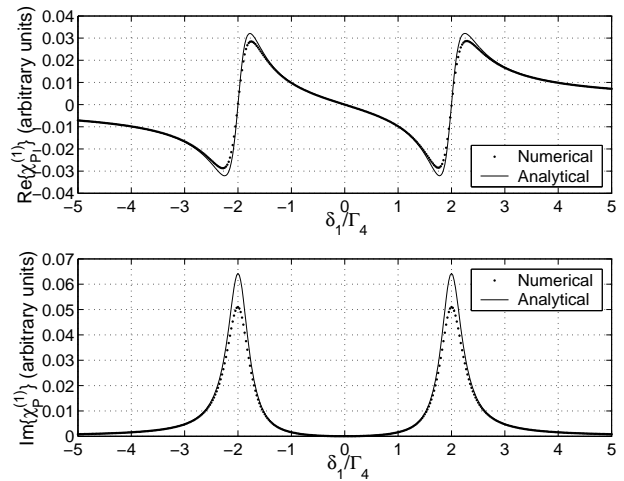}
\caption{Comparison of the numerical solution (dotted line) of the OBE with the analytical prediction
of Eq.~(\protect\ref{eq:probelinasym}) (full line) for the real part (above) and imaginary part (below) of the linear probe susceptibility versus
the normalized probe detuning $\delta_{1}/\Gamma_4$. The parameter used are the following:  $\Gamma_2^{AV}=\Gamma_{2} = 36$ MHz,
$\Gamma_4^{AV}=\Gamma_{4}= 38$ MHz, $\delta_{2} = \delta_{3} = \delta_{4} = 0$,  $\forall$ i,j $\gamma_{ij}=\Gamma_3^{AV}=\Gamma_5^{AV} =
10^{-4}\Gamma_{4}$, $\Omega_{1}=0.08\Gamma_{4}$, $\Omega_{2} = 2\Gamma_{4}$, $\Omega_{3} = 0.04\Gamma_{4}$,
$\Omega_{4} = \Gamma_{4}$.}
\label{fig3}
\end{center}
\end{figure}

\begin{figure}[t]
\begin{center}
\includegraphics[scale=0.45]{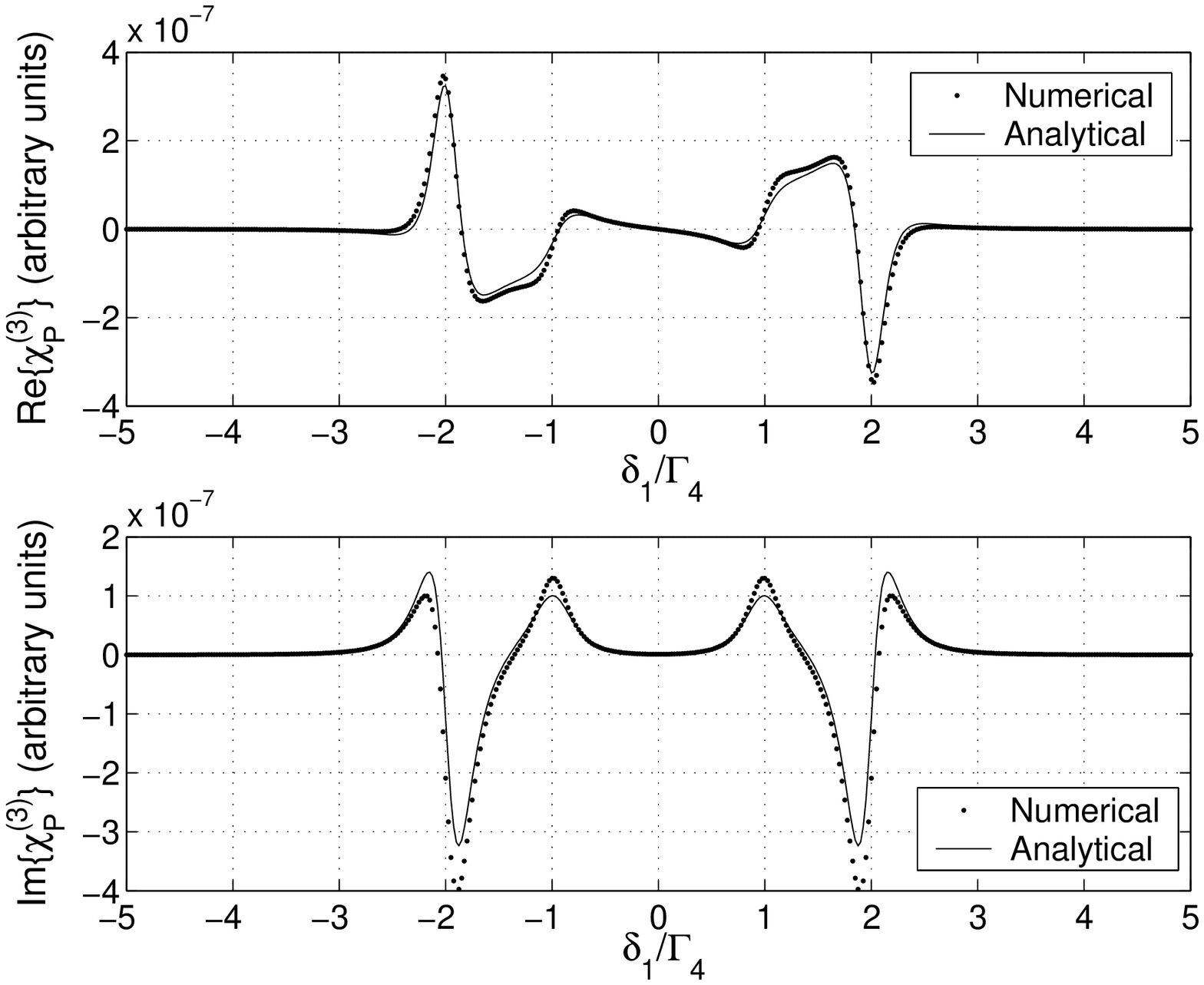}
\caption{Comparison of the numerical solution (dotted line) of the OBE with the analytical prediction
of Eq.~(\protect\ref{eq:suscgennonlin}b) (full line) for the real part (above) and imaginary part (below) of the probe cross-Kerr susceptibility
versus the normalized probe detuning $\delta_{1}/\Gamma_4$.
To reduce as much as possible the influence of the self-Kerr susceptibility we have considered a probe Rabi frequency
$\Omega_{1}$ much smaller than that of the trigger field.
Parameters are: $\Gamma_2^{AV}=\Gamma_{2} = 36$ MHz,
$\Gamma_4^{AV}=\Gamma_{4}= 38$ MHz, $\delta_{2} = \delta_{3} = \delta_{4} = 0$,  $\forall$ i,j $\gamma_{ij}=\Gamma_3^{AV}=\Gamma_5^{AV} =
10^{-4}\Gamma_{4}$, $\Omega_{1}=0.004\Gamma_{4}$, $\Omega_{2} = 2\Gamma_{4}$, $\Omega_{3} = 0.04\Gamma_{4}$,
$\Omega_{4} = \Gamma_{4}$.}
\label{fig4}
\end{center}
\end{figure}

\begin{figure}[t]
\begin{center}
\includegraphics[scale=0.45]{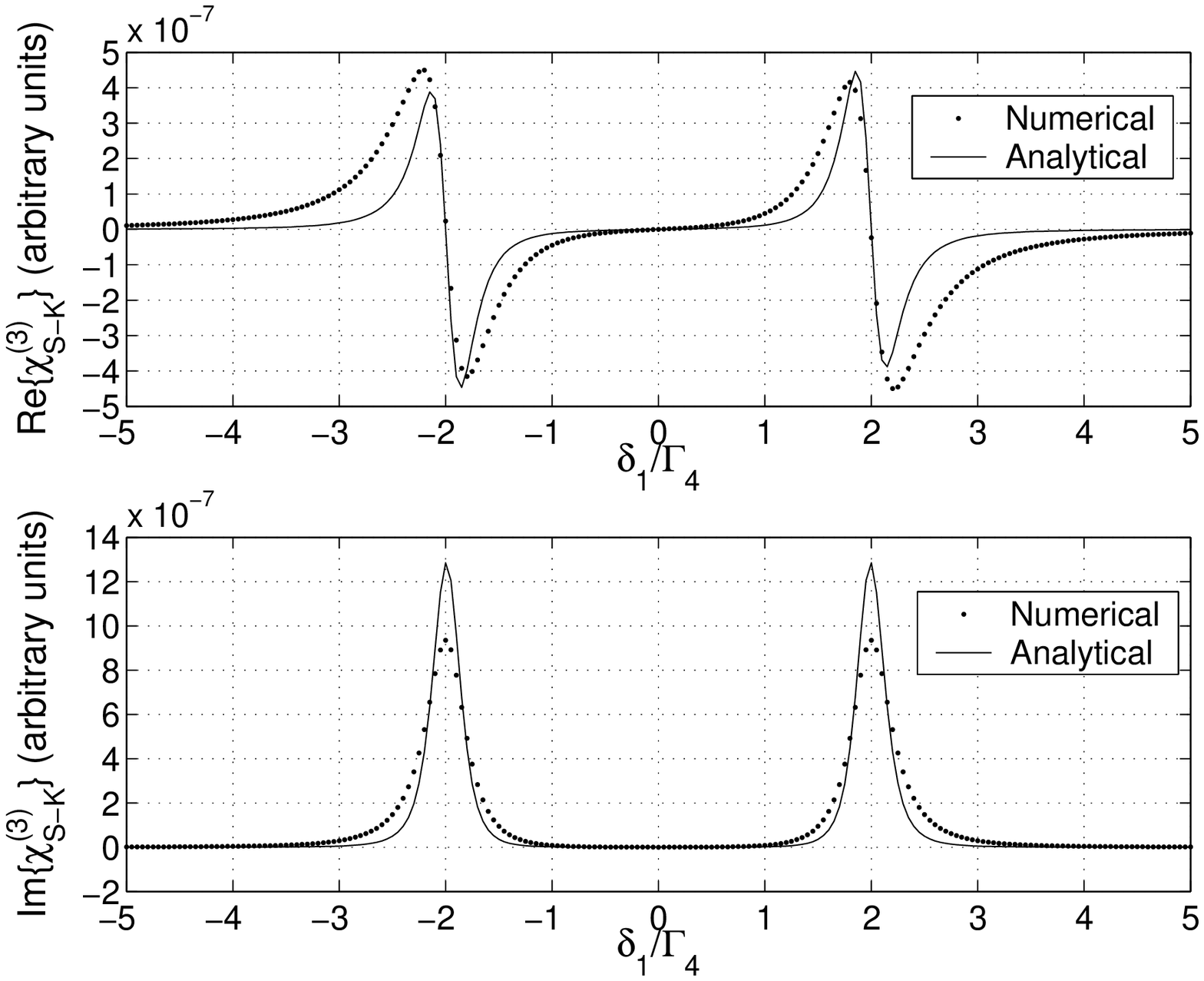}
\caption{Comparison of the numerical solution (dotted line) of the OBE with the analytical prediction
of Eq.~(\protect\ref{eq:suscgennonlin}a) (full line) for the real part (above) and imaginary part (below) of the probe self-Kerr susceptibility
versus the normalized probe detuning $\delta_{1}/\Gamma_4$.
To reduce as much as possible the influence of the cross-Kerr susceptibility we have considered a trigger Rabi frequency
$\Omega_{3}$ much smaller than that of the probe field.
Parameters are: $\Gamma_2^{AV}=\Gamma_{2} = 36$ MHz,
$\Gamma_4^{AV}=\Gamma_{4}= 38$ MHz, $\delta_{2} = \delta_{3} = \delta_{4} = 0$,  $\forall$ i,j $\gamma_{ij}=\Gamma_3^{AV}=\Gamma_5^{AV} =
10^{-4}\Gamma_{4}$, $\Omega_{1}=0.5\Gamma_{4}$, $\Omega_{2} = 2\Gamma_{4}$, $\Omega_{3} = 0.005\Gamma_{4}$,
$\Omega_{4} = \Gamma_{4}$.}
\label{fig4bis}
\end{center}
\end{figure}

\begin{figure}[t]
\begin{center}
\includegraphics[scale=0.45]{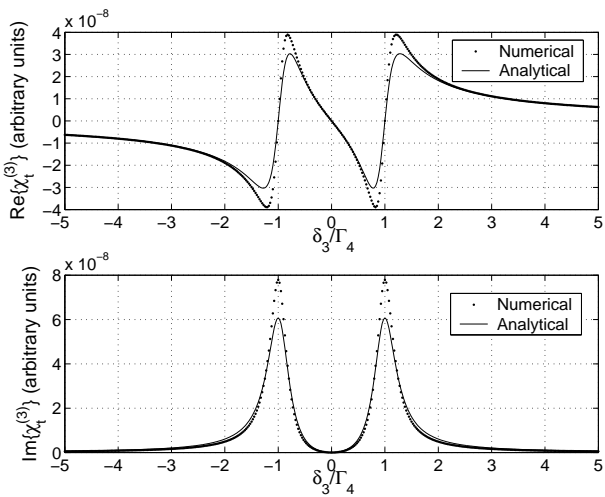}
\caption{Comparison of the numerical solution (dotted line) of the OBE with the analytical prediction
of Eq.~(\protect\ref{eq:suscgennonlin}c) (full line) for the real part (above) and imaginary part (below) of the
trigger cross-Kerr susceptibility as a function of the normalized
trigger's detuning $\delta_3/\Gamma_4$.
Parameters are similar to those of Fig.~\ref{fig3}, $\Gamma_2^{AV}=\Gamma_{2} = 36$ MHz,
$\Gamma_4^{AV}=\Gamma_{4}= 38$ MHz, $\delta_{1} = \delta_{2} = \delta_{4} = 0$,  $\forall$ i,j $\gamma_{ij}=\Gamma_3^{AV}=\Gamma_5^{AV} =
10^{-4}\Gamma_{4}$, $\Omega_{1}=0.08\Gamma_{4}$, $\Omega_{2} = 2\Gamma_{4}$, $\Omega_{3} = 0.04\Gamma_{4}$,
$\Omega_{4} = \Gamma_{4}$.} \label{fig5}
\end{center}
\end{figure}

We then consider the numerical solution of the OBE and we compare it with the analytical treatment based on the
AV approach presented above. The numerical calculations are performed in the range of parameters corresponding to EIT,
i.e., $|\Omega_1|,|\Omega_2| \ll |\Omega_3|, |\Omega_4|$ and we stay near two-photon resonance for both the probe and the trigger field.
In Figs.~\ref{fig3}-\ref{fig5} we compare the analytical solutions of Eq.~(\ref{eq:probelinasym}) and Eqs.~(\ref{eq:suscgennonlin}) with the
numerical solution of the complete set of Bloch equations given in the Appendix A.
From these plots it is evident that our analytical treatment works satisfactorily well, except for a small interval of values of
the detuning, corresponding to the maximum probe (or trigger) absorption. In such a case, the detunings match the Rabi frequencies of the
two pumps, and the probe (or trigger) field is in resonance with a single atomic transition. The atoms are significantly pumped to the excited levels
and the population assumption of Eq.~(\ref{eq:adiabass}) is not fulfilled. In fact, the discrepancy between the exact numerical solution of the OBE
and the AV approach is strictly related to the atomic population
out of level $|1\rangle$ which, in the case of Fig.~\ref{fig3}, is about $14\%$ of the total population.

Figs.~\ref{fig3}-\ref{fig5} refer to a situation with small dephasing rates ($\forall$ i,j, $\gamma_{ij}=\Gamma_3^{AV}=\Gamma_5^{AV}=10^{-4} \Gamma_4 \sim $ few kHz)
which are
typical for not too dense gases. For larger values of the dephasing rates (some tens of kHz), we have seen that
the analytical prediction of the AV approach of the preceding subsection
starts to depart from the exact solution of the OBE.

\subsection{Group velocity matching}\label{sec:vgmatch}

The propagation equation for the slowly varying electric field amplitudes
$\varepsilon_{i}(z,t)$, $i= P,T$, defined as
$$
{\mathcal E}_{i}(z,t)=\varepsilon_{i}(z,t)
\exp\left\{ik_{i}z-i\omega_{i}t\right\}+c.c.\;\;i=P,T,
$$ is given by
\begin{equation}
\left(\frac{\partial}{\partial z}+\frac{1}{v_{g}^{i}}
\frac{\partial}{\partial t}\right)\varepsilon_{i}(z,t)=
i\frac{k_{i}}{2}\chi_{i}(z,t)\varepsilon_{i}(z,t), \;\;i=P,T,
\label{propa}
\end{equation}
where $v_{g}^{i}$ is the group velocity, generally defined as
$v_g^{i} = c/(1+n_g^{i})$, with $c$ the speed of light in
vacuum and
\begin{equation}
n_g^{i} = \frac{1}{2} {\rm Re}[\chi_{i}] + \frac{\omega_i}{2} \left(\frac{\partial
{\rm Re}[\chi_{i}]}{\partial \omega} \right)_{\omega_i} \label{eq:ng}
\end{equation}
the group index, $\omega_i$ being the frequency of field $i$.
The solution of Eq.~(\ref{propa}) is
\begin{equation}
\varepsilon_{i}(z,t)=\varepsilon_{i}(0,t-\frac{z}{v_{g}^{i}})
\exp\left\{
i\frac{k_{i}}{2}\int_{0}^{z}dz'\chi_{i}(z',t)\right\},
\label{propa2}
\end{equation}
so that, using Eqs.~(\ref{eq:suscgen}), the nonlinear cross-phase shift for the two fields of interest is given by
\begin{subequations}
\label{crosspha}
\begin{eqnarray}
&& \phi_{P}^{ck}=  \frac{\omega_1}{2c}\int_{0}^{l}dz \mathrm{Re}[\chi_{P}^{3,ck}]\left|\varepsilon_{T}(z,t)\right|^2,\\
&& \phi_{T}^{ck}=  \frac{\omega_3}{2c}\int_{0}^{l}dz \mathrm{Re}[\chi_{T}^{3,ck}]\left|\varepsilon_{P}(z,t)\right|^2,
\end{eqnarray}
\end{subequations}
where $l$ is the length of the atomic medium. These nonlinear cross-phase shifts are of fundamental importance
also for quantum information processing applications. In fact, the CPS of Eq.~(\ref{eq:def_cps}) is determined
only by these cross-Kerr contributions to the total phase shift, because the linear and self-Kerr contributions cancel out,
as shown in Refs.~\cite{nostro,tripod}.

For Gaussian probe and trigger pulses of time durations $\tau_P$ and $\tau_T$, and with peak Rabi frequencies $\Omega_P^{peak}$ and $\Omega_T^{peak}$
respectively, the nonlinear cross-phase shifts can be written as (see also Refs.~\cite{nostro,tripod})
\begin{subequations}
\label{crosspha2}
\begin{eqnarray}
 && \phi_{P}^{ck} = \frac{\omega_1 l}{4c}
\frac{\sqrt{\pi}\hbar^2|\Omega_T^{peak}|^2}{|\mu_{34}|^2}\,
\frac{\rm{erf}[\zeta_P]}{\zeta_P} \, \mathrm{Re}[\chi_P^{3,ck}],\\
&& \phi_{T}^{ck} = \frac{\omega_3 l}{4c}
\frac{\sqrt{\pi}\hbar^2|\Omega_P^{peak}|^2}{|\mu_{12}|^2}\,
\frac{\rm{erf}[\zeta_T]}{\zeta_T} \, \mathrm{Re}[\chi_T^{3,ck}],
\end{eqnarray}
\end{subequations}
where $\zeta_P = (1-v_g^P/v_g^T)\sqrt{2}l/v_g^P\tau_T$ and $\zeta_T$ is obtained from $\zeta_P$ upon interchanging the indices $P
\leftrightarrow T$.
Large nonlinear cross-phase shifts take place for
appreciably large values of the two cross-Kerr susceptibilities real parts, and especially when probe and trigger velocities become
\emph{equal}, i.e., when
$\zeta_{P,T} \rightarrow 0$, in which case the erf[$\zeta$]/$\zeta$ reaches the maximum value $2/\sqrt{\pi}$.
In this limit the cross-phase phase shifts linearly increase with the length of the atomic medium $l$.
This explains why achieving group velocity matching, $v_g^P=v_g^T$, is of fundamental importance.
Moreover group velocities become small for large group indices and this condition can be achieved within the EIT transparency window,
where Re[$\chi $] vanishes, and the group velocity is strongly reduced due to a large dispersion gradient $\partial
{\rm Re}[\chi]/\partial \omega$.

\begin{figure}[t]
\begin{center}
\includegraphics[scale=0.45]{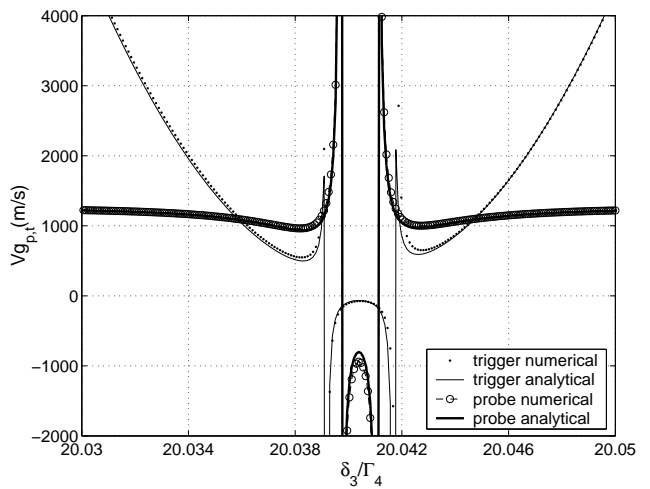}
\caption{Group velocity of the probe and trigger pulses versus the normalized trigger detuning $\delta_3/\Gamma_4$.
Full lines denote the analytical predictions of Eqs.~(\protect\ref{eq:vgAsym}) (the thick line refers to the probe and the thin line to the trigger).
Circles and dots refer to the numerical solution of the OBE for the probe and trigger group velocity, respectively.
This figure shows how it is possible to obtain group velocity matching in the asymmetric \emph{M}-scheme: two different values
of $\delta_3$ exist for which $v_{g}^{P} = v_{g}^{T} \simeq 1000$ m/s.
The parameters are those of the $D_{1}$ and $D_{2}$ line in the $^{87}$Rb spectrum: $\Gamma_2^{AV}=\Gamma_{2} \simeq 36$ MHz, $\Gamma_4^{AV}=\Gamma_{4}\simeq 38$ MHz,
$\delta_{1}=\delta_{2}=0$, $\delta_{4} \simeq \delta_{3} \simeq 20\Gamma_{4}$, $\Omega_{1}=0.08\Gamma_{4}$,
$\Omega_{2} = 2\Gamma_{4}$, $\Omega_{3} = 0.04\Gamma_{4}$, $\Omega_{4} = \Gamma_{4}$,
$\forall$ i,j, $\gamma_{ij}=\Gamma_3^{AV}=\Gamma_5^{AV}=10^{-4}\Gamma_{4}$, $N/V=3.0\cdot10^{13}$ cm$^{-3}$.}
  \label{vg_prl}
\end{center}
\end{figure}

Let us see how small and equal probe and trigger group velocities can be obtained.
We consider the approximate analytical expressions for the susceptibilities of Eqs.~(\ref{eq:suscgen})-(\ref{eq:suscgennonlin})
derived above within the AV approach, and which we have seen to work very well in the EIT regime.
Assuming to stay at the center of the transparency window for the probe
($\delta_{12}=0$) where the dispersion gradient is maximum, and neglecting dephasing rates $\Gamma_3^{AV}$ and
$\Gamma_5^{AV}$, which are typically much smaller than all the other parameters, one gets
\begin{subequations}
\label{eq:dchip}
\begin{eqnarray}
n_g^{P} &\simeq & \frac{N}{V} \frac{|\mu_{12}|^{2}\omega_1} {2\hbar\epsilon_{0}|\Omega_{2}|^{2}}(1+|\Omega_{3}|^{2}\beta ),\\
n_g^{T} &\simeq & \frac{N}{V} \frac{|\mu_{34}|^{2}\omega_3} {2\hbar\epsilon_{0}|\Omega_{2}|^{2}}|\Omega_{1}|^{2}\beta,
\end{eqnarray}
\end{subequations}
where \cite{nostro}
\begin{equation}\label{eq:betab}
\beta = \frac{\left(\delta_{14}^{2}+|\Omega_{4}|^{2}\right) \left[\left(\delta_{13}\delta_{14}-|\Omega_{4}|^{2}
 \right)^{2} -\delta_{14}^2\left(\Gamma_4^{AV}/2\right)^2\right]}{\left[ \left( \delta_{13}\delta_{14} -|\Omega_{4}|^{2}\right)^{2} +
 \delta_{14}^2 \left(\Gamma_4^{AV}/2\right)^2\right]^{2}}.
\end{equation}
In the EIT situation we are considering it is $n_g^P, n_g^T \gg 1$, so that, using Eqs.~(\ref{eq:dchip}),
\begin{subequations}
\label{eq:vgAsym}
\begin{eqnarray}
&& v_{g}^{P} \simeq \frac{c}{n_g^{P}}\simeq \frac{2\hbar\epsilon_{0}c|\Omega_{2}|^{2}} {(N/V)|\mu_{12}|^{2}\omega_1(1+|\Omega_{3}|^{2}\beta )}, \\
&& v_{g}^{T} \simeq \frac{c}{n_g^{T}} \simeq \frac{2\hbar\epsilon_{0}c|\Omega_{2}|^{2}} {(N/V)|\mu_{34}|^{2}\omega_3|\Omega_{1}|^{2}\beta}.
\end{eqnarray}
\end{subequations}
As expected, the asymmetric \emph{M}-scheme does not yield equal slow down of both trigger and probe pulse
automatically as, for example, the scheme of Petrosyan and Kurizki~\cite{kuri} does. In fact, the two expressions of the group velocities are generally different.
Nonetheless, Eqs.~(\ref{eq:vgAsym}) show that group velocity matching is always achievable by properly adjusting the parameter $\beta$,
which means adjusting the tuner intensity $|\Omega_4|^2$ and the composite detuning $\delta_{14}$. This shows that the present
asymmetric \emph{M}-scheme can be seen as a modified version of the \emph{N}-scheme of Ref.~\cite{ima}, in which the tuner pump field is added
just in order to ``tune'' the group velocity of the trigger pulse so to make it equal to that of the probe. The possibility to achieve
group velocity matching is shown in Fig.~\ref{vg_prl}, where both the numerical result derived form the OBE and the approximate analytical expressions
of Eqs.~(\ref{eq:vgAsym}) are plotted versus the trigger detuning $\delta_3$. Two different values
of $\delta_3$ exist for which $v_{g}^{P} = v_{g}^{T} \simeq 1000$ m/s (see Fig.~\ref{vg_prl}). Parameter values here correspond to typical values
for a cell of $^{87}$Rb atoms, i.e., $\Gamma_2^{AV}=\Gamma_2 \simeq 36$ MHz,  $\Gamma_4^{AV}=\Gamma_4\simeq 38$ MHz,
$N/V \simeq 3 \times 10^{13}$ cm$^{-3}$, $\delta_{1}=\delta_{2}=0$,
$\delta_{4} \simeq \delta_{3} \simeq 20\Gamma_{4}$, $\Omega_{1}=  0.08\Gamma_{4}$, $\Omega_{2} = 2\Gamma_{4}$, $\Omega_{3} =
0.04\Gamma_{4}$, $\Omega_{4}= \Gamma_{4}$, $\forall$ i,j, $\gamma_{ij}=\Gamma_3^{AV}=\Gamma_5^{AV}=10^{-4}\Gamma_{4}$. Moreover Fig.~\ref{vg_prl} clearly shows that the
simple expressions of Eqs.~(\ref{eq:vgAsym}) well reproduce the exact numerical solution of the OBE.

\subsection{Pulse propagation}\label{sec:pulseprop}

In previous Section, we have addressed the problem of group velocity matching between probe and trigger fields in the asymmetric M-scheme. It should be emphasized that the analysis and the results presented there are strictly valid for the continuous-wave (cw) fields. We would now address that same problem but with the pulsed probe and trigger fields in mind. At the first look, Eqs.~(\ref{eq:vgAsym}) appear to suggest that the group velocity matching would not be possible in the pulsed regime. As the group velocity of the trigger pulse is inversely proportional to the square of the probe pulse, trigger suffers anomalous dispersion, i.e. in the presence of a pulsed probe, the trigger pulse will get distorted, splitting into several components, each having a different group velocity. 

\begin{figure*}[t]
\begin{center}
\includegraphics[scale=0.8]{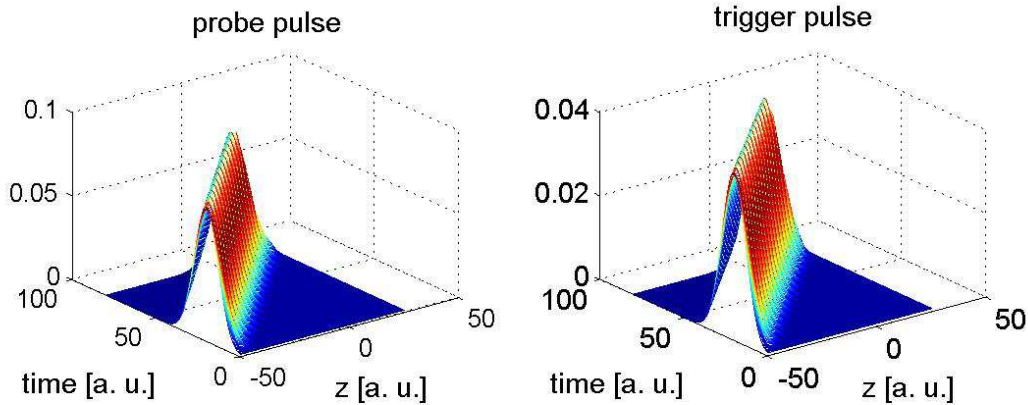}
\caption{(Color online) Propagation of probe and trigger pulses through the asymmetric M medium. Pulses are taken to be Gaussian at time $t=0$ and are sufficiently long $(\tau_i > 1/\Delta\omega_{tr}^i)$, $i=P, T$. Units are arbitrary, with $c=1$.}
\label{fig:longpulses}
\end{center}
\end{figure*}

\begin{figure*}[t]
\begin{center}
\includegraphics[scale=0.8]{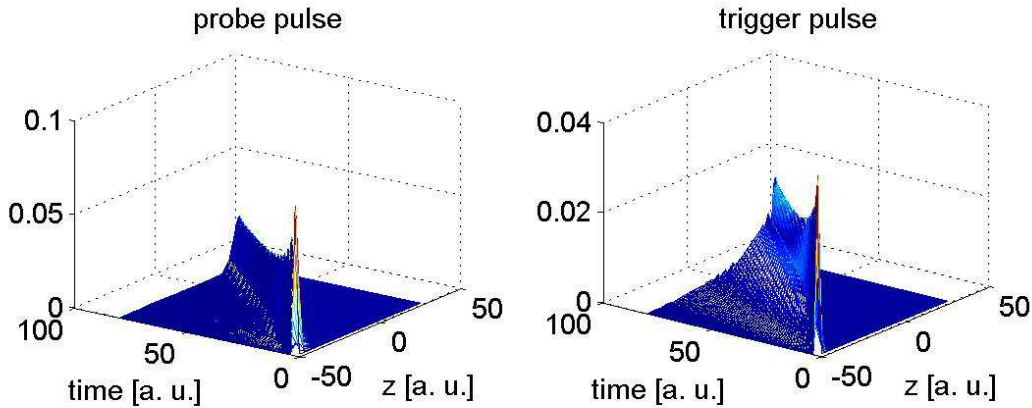}
\caption{(Color online) Propagation of probe and trigger pulses through the asymmetric M medium. Pulses are taken to be Gaussian at time $t=0$ and short $(\tau_i < 1/\Delta\omega_{tr}^i)$, $i=P, T$. Units are arbitrary, with $c=1$.}
\label{fig:shortpulses}
\end{center}
\end{figure*}

It will be shown in this Section that the above conclusion is an artifact of approximations made to obtain a closed and compact expression for group velocities. In particular, the pulse propagation in this approximation is described by Eqs.~(\ref{propa}), with group velocities $v_g^i$ given by Eqs.~(\ref{eq:vgAsym}) and nonlinear susceptibilities $\chi_i$ being those of Eqs.~(\ref{eq:suscgennonlin}). This is equivalent to the adiabatic elimination of the atomic degrees of freedom. Such adiabatic elimination, strictly speaking, is not valid in the parameter regime explored in this paper: strong nonlinear interaction between probe and trigger pulses suggests that the contribution of the atomic medium is far from being adiabatic. Also, it should be noted that the dephasing processes have been neglected in the derivation of $v_g^i$. For the adiabatic case, the above conclusion is correct: the trigger pulse suffers anomalous dispersion and its group velocity becomes singular towards the edges of a probe pulse. However, pulse propagation through the asymmetric M-system do not follow such a simple approximate evolution. Full propagation problem must then be solved which includes adding the time-dependent equations for the pulses 
\begin{eqnarray}
\left(\frac{\partial}{\partial z}+\frac{1}{c}
\frac{\partial}{\partial t}\right)\varepsilon_{i}(z,t)=
i\frac{k_{i}}{2}\frac{N\mu_i}{V\epsilon_0}\sigma_{i}(z,t), \;\;i=P,T,
\label{eq:pulsepropfull}
\end{eqnarray}
to the OBEs [Eqs.~(\ref{eq:blochpopAsym},~\ref{eq:blochcohAsym})], and numerically solving the resulting system of equations. In the above equation, it is understood that $\sigma_P = \sigma_{12}$ and $\sigma_T = \sigma_{34}$ in the notation of Eqs.~(\ref{eq:blochcohAsym}). 

Results are shown in Figs.~\ref{fig:longpulses} and~\ref{fig:shortpulses} for the same set of parameters that yields group velocities matching in Fig.~\ref{vg_prl}. Vertical axes have been scaled appropriately to obtain Rabi frequencies $\Omega_i$. Two operating regimes could be identified, long-pulse regime and short-pulse regime, where `long' and `short' denotes the pulses' length in time. This length is compared to the inverse width of the transparency window. Long pulses fit well into the transparency window, while short pulses do not. Fig.~\ref{fig:longpulses} shows the results of our simulation for the initially identical long Gaussian pulses. It is clear that the pulses propagate undistorted with the equal group velocities. Tiny amplitude decay is present due to the small imaginary part of the nonlinear susceptibility. 

Short pulses (Fig.~\ref{fig:shortpulses}) however, show distortion. Probe pulse distortion comes from the absorption, as the pulse spreads outside of the transparency window. Trigger pulse shows the same absorption effect, but moreover it also splits into several components which then continue to propagate with a different group velocities each. Note that the singularity present in the adiabatic approach is not present here. This is due to the fact that the dephasing, neglected in the adiabatic treatment, effectively \textit{regularizes} the equations, removing the singularity.

It is also noted that in the long-pulse regime, both of the pulses propagate virtually undistorted, with a group velocity uniform across each of the pulses. Our simulations suggest that the approximate Eqs.~(\ref{eq:vgAsym}) are valid, as long as the Rabi frequencies there are considered to be taken at the \textit{peak} of the pulse, i.e. $\Omega_i \rightarrow \Omega_i^{peak}$.

\section{The Symmetric M--scheme}\label{sec:symmM}

\begin{figure}[t]
\begin{center}
\includegraphics[scale=0.8]{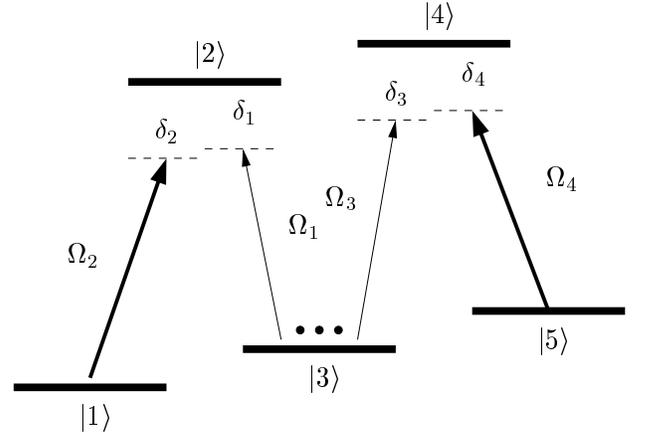}
\caption{Symmetric \emph{M} scheme. The probe and the trigger fields, with Rabi frequencies $\Omega_{1}$ and $\Omega_{3}$ respectively,
together with the stronger pump fields, the coupler and the tuner (with Rabi frequencies $\Omega_{2}$ and $\Omega_{4}$, respectively)
drive the corresponding transitions. All the atoms are assumed to be in state $|3\rangle $ and
the detunings are defined in Eqs.~(\protect\ref{eq:det2}).}
\label{fig:SMS}
\end{center}
\end{figure}

In this section we analyze the symmetric M--scheme, schematically shown in Fig.~\ref{fig:SMS}. The initial conditions and the
configuration of the fields are slightly different from those of the asymmetric case of Sec.~\ref{sec:assym}. The same five levels
could be used, but all the atoms are now initially prepared in level $|3\rangle$ (see Fig.~\ref{fig:SMS}). Moreover, the role of the probe and of the coupler
fields are exchanged, i.e., now the probe field (still with Rabi frequency $\Omega_1$ and central frequency $\omega_1$) couples
levels $|2\rangle $ and $|3\rangle $, while the coupler (still with Rabi frequency $\Omega_2$ and central frequency $\omega_2$)
induces transitions between levels $|1\rangle $ and $|2\rangle $. The role of trigger and tuner fields remains unchanged. In such a way, the scheme becomes
symmetric for probe and trigger, and the two fields experience \emph{exactly} the same dynamics whenever the corresponding parameters are made equal, i.e.,
when the Rabi frequencies are correspondingly equal ($\Omega_1=\Omega_3$, $\Omega_2=\Omega_4$), as well as the detunings,
($\delta_1=\delta_3$, $\delta_2=\delta_4$), which
are now defined similarly to those of the asymmetric \emph{M} scheme (see Eqs.~(\ref{eq:det})) except for probe-coupler exchange,
i.e.,
\begin{subequations}
\label{eq:det2}
\begin{eqnarray}
E_{2}-E_{1}&=&\hbar\omega_{2}+\hbar\delta_{2},\\
E_{2}-E_{3}&=&\hbar\omega_{1}+\hbar\delta_{1}, \\
E_{4}-E_{3}&=&\hbar\omega_{3}+\hbar\delta_{3}, \\
E_{4}-E_{5}&=&\hbar\omega_{4}+\hbar\delta_{4}.
\end{eqnarray}
\end{subequations}
In this way, the scheme can be seen again as formed by two adjacent $\Lambda$, one for the probe and one for the trigger,
now however symmetrically placed with respect to state $|3\rangle $.
As we have done for the asymmetric \emph{M} scheme, we assume to stay close to the two-photon resonance conditions, $\delta_1 \simeq \delta_2$ and
$\delta_3 \simeq \delta_4$, and moreover that $|\Omega_1 | \ll |\Omega_2|$, and $|\Omega_3 |\ll |\Omega_4|$, so that both probe and trigger will experience EIT.
As we have seen above, a large XPM is obtained when the group velocities are equal~\cite{harhau,luima,nostro,kuri,tripod,noipra},
and the advantage of the present symmetric M--scheme
is that group velocity matching is automatically achieved once that the scheme is exactly symmetric between probe and trigger.

The Hamiltonian of the system is
\begin{eqnarray}\label{eq:ham1s}
H_S &=& \sum_{i}^{5}E_{i}|i\rangle\langle
i|+\hbar\left(\Omega_{1}e^{-i\omega_{1}t}|2\rangle\langle 3|+\Omega_{2}e^{-i\omega_{2}t}|2\rangle\langle 1|\right. \nonumber\\
&&+\left. \Omega_{3}e^{-i\omega_{3}t}|4\rangle\langle3|+\Omega_{4}e^{-i\omega_{4}t}|4\rangle\langle5|+h.c.\right).
\end{eqnarray}\\
Moving to the interaction picture with respect to the following
free Hamiltonian
\begin{eqnarray}\label{eq:hamfrees}
H_{0}' &=&
E_{3}|3\rangle\langle 3|+(E_{2}-\hbar\delta_{1})|2\rangle\langle2|+(E_{1}-\hbar\delta_{12})|1\rangle\langle 1| \nonumber \\
&& +(E_{4}-\hbar\delta_{3})|4\rangle\langle 4|+(E_{5}-\hbar\delta_{34})|5\rangle\langle5| ,
\end{eqnarray}\\
where
\begin{subequations}
\label{eq:deltas}
\begin{eqnarray}
\delta_{12} &=& \delta_{1}-\delta_{2},\\
\delta_{34} &=& \delta_{3}-\delta_{4},
\end{eqnarray}
\end{subequations}
we get the following effective Hamiltonian
\begin{eqnarray}\label{eq:hameffs}
&& H_{eff}^{S} =\hbar\delta_{1}|2\rangle\langle2|+\hbar\delta_{12}|1\rangle\langle 1|+\hbar\delta_{3}|4\rangle\langle 4|+\hbar\delta_{34}|5\rangle\langle5| \nonumber\\
&&+\hbar \Omega_{1}|2\rangle\langle 3|+\hbar \Omega_{2}|2\rangle\langle 1|+\hbar \Omega_{3}|4\rangle\langle3|+\hbar \Omega_{4}|4\rangle\langle5| \nonumber\\
&&+\hbar \Omega_{1}^{\star}|3\rangle\langle 2|+\hbar \Omega_{2}^{\star}|1\rangle\langle 2|+\hbar \Omega_{3}^{\star}|3\rangle\langle4|+\hbar \Omega_{4}^{\star}
|5\rangle\langle4|.
\end{eqnarray}

\subsection{Amplitude Variables Approach}

We first study the system dynamics by means of the AV approach, in which
the state of the atom is described by the wave-function of Eq.~(\ref{eq:state}), whose time evolution is determined by the
Hamiltonian of Eq.~(\ref{eq:hameffs}), supplemented with phenomenological decay rates $\Gamma_{i}^{AV}$ for each atomic
level $|i\rangle $. The corresponding evolution equations for the amplitudes $b_i(t)$ are
\begin{subequations}
\label{eq:AVsymm}
\begin{eqnarray}
\dot{b}_{1}(t) &= &-\imath d_1 b_{1}(t) - \imath\Omega^{\star}_{2}b_{2}(t),\\
\dot{b}_{2}(t) &=& -\imath d_2 b_{2}(t) - \imath\Omega_{2}b_{1}(t) - \imath\Omega_{1}b_{3}(t), \\
\dot{b}_{3}(t) &=& -\imath d_3 b_{3} - \imath\Omega^{\star}_{1}b_{2}(t) - \imath\Omega^{\star}_{3}b_{4}(t), \\
\dot{b}_{4}(t) &=& -\imath d_4 b_{4}(t) - \imath\Omega_{3}b_{3}(t) - \imath\Omega_{4}b_{5}(t), \\
\dot{b}_{5}(t) &=& -\imath d_5 b_{5}(t) - \imath\Omega^{\star}_{4}b_{4}(t),
\end{eqnarray}
\end{subequations}
where, similarly to what we have done for the asymmetric case, we have defined
\begin{subequations}
\label{eq:complexdet}
\begin{eqnarray}
d_1 &=& \delta_{12}-\imath\Gamma_{1}^{AV}/2,\\
d_2 &=& \delta_{1}-\imath \Gamma_{2}^{AV}/2, \\
d_3 &=& -\imath\Gamma_{3}^{AV}/2, \\
d_4 &=& \delta_{3}-\imath \Gamma_{4}^{AV}/2, \\
d_5 &=& \delta_{34}-\imath \Gamma_{5}^{AV}/2.
\end{eqnarray}
\end{subequations}
Since we choose again $|\Omega_1/\Omega_2| \ll 1$ and $|\Omega_3/\Omega_4| \ll 1$, it is reasonable
to assume that the atomic population remains in the initial state $|3\rangle$ to a good approximation
\begin{equation}\label{eq:popassumption}
b_{3}^{ss}\sim 1.
\end{equation}
The set of equations~(\ref{eq:AVsymm}) is then solved in the steady-state. In order to get a consistent expression for the nonlinear susceptibilities one has to
consider higher order contributions to Eq.~(\ref{eq:popassumption}), which is obtained by imposing the normalization of the atomic wave-function
of Eq.~(\ref{eq:state}) at second order in $|\Omega_1/\Omega_2|$ and $|\Omega_3/\Omega_4|$. One gets the following expression for the steady state amplitudes
\begin{subequations}
\label{eq:amplisolsym}
\begin{eqnarray}
&& b_{3}^{ss}=1-\frac{\left|\Omega_{1}\right|^2 \left[\left|d_{1}\right|^2+\left|\Omega_{2}\right|^2\right]}{2\left|d_{1}d_{2}-|\Omega_{2}|^2\right|^2}
\nonumber \\
&& -\frac{\left|\Omega_{3}\right|^2 \left[\left|d_{5}\right|^2+\left|\Omega_{4}\right|^2\right]}{2\left|d_{4}d_{5}-|\Omega_{4}|^2\right|^2},\\
&& b_{2}^{ss}=-\frac{\Omega_{1}d_{1}}{d_{1}d_{2}-|\Omega_{2}|^2}b_{3}^{ss},\\
&& b_{4}^{ss}=-\frac{\Omega_{3}d_{5}}{d_{5}d_{4}-|\Omega_{4}|^2}b_{3}^{ss},\\
&& b_{1}^{ss}=\frac{\Omega_{1}\Omega^{\star}_{2}}{d_{1}d_{2}-|\Omega_{2}|^2}b_{3}^{ss} ,\\
&& b_{5}^{ss}=\frac{\Omega_{3}\Omega^{\star}_{4}}{d_{5}d_{4}-|\Omega_{4}|^2}b_{3}^{ss}.
\end{eqnarray}
\end{subequations}
These results can be used to determine the probe and trigger susceptibilities, which are now defined as (see Eqs.~(\ref{eq:avchi}))
\begin{subequations}
\label{eq:avchisym}
\begin{eqnarray}
\chi_{P} &=& \frac{N \mu_{32}}{V\varepsilon_{0}{\mathcal E}_1}b_{2}^{ss}b_{3}^{ss,\star}=-\frac{N |\mu_{32}|^2}{V\hbar \varepsilon_{0}\Omega_1}
b_{2}^{ss}b_{3}^{ss,\star}, \\
\chi_{T} &=& \frac{N \mu_{34}}{V\varepsilon_{0}{\mathcal E}_3}b_{4}^{ss}b_{3}^{ss,\star}=-\frac{N |\mu_{34}|^2}{V\hbar \varepsilon_{0}\Omega_3}
b_{4}^{ss}b_{3}^{ss,\star}.
\end{eqnarray}
\end{subequations}
Inserting Eqs.~(\ref{eq:amplisolsym}) into Eqs.~(\ref{eq:avchisym}) and expanding
in series at the lowest orders in the probe and trigger electric fields, ${\mathcal E}_{1}$ and ${\mathcal E}_{3}$
respectively, one gets
\begin{subequations}
\label{eq:suscgensym}
\begin{eqnarray}
\chi_{P}&\simeq&\chi^{(1)}_{P}+\chi^{(3,sk)}_{P}|{\mathcal E}_{1}|^{2}+\chi^{(3,ck)}_{P}|{\mathcal E}_{3}|^{2},\\
\chi_{T}&\simeq&\chi^{(1)}_{T}+\chi^{(3,sk)}_{T}|{\mathcal E}_{3}|^{2}+\chi^{(3,ck)}_{T}|{\mathcal E}_{1}|^{2},
\end{eqnarray}
\end{subequations}
where we have again distinguished the third-order self-Kerr susceptibilities $\chi^{(3,sk)}_{P,T}$ from the
third-order cross-Kerr susceptibilities $\chi^{(3,ck)}_{P,T}$. Using Eqs.~(\ref{eq:amplisolsym}) and the definitions of
Eqs.~(\ref{eq:complexdet}), we get the following expressions for the linear susceptibilities,
\begin{subequations}
\label{eq:probelinsym}
\begin{eqnarray}
&& \chi^{(1)}_{P}=\frac{N |\mu_{32}|^2}{V\hbar \varepsilon_{0}}\frac{\delta_{12}-i\Gamma_1^{AV}/2}{\left(\delta_{1}-i\Gamma_2^{AV}/2\right)
\left(\delta_{12}-i\Gamma_1^{AV}/2\right)-|\Omega_{2}|^{2}}\\
&& \chi^{(1)}_{T}=\frac{N |\mu_{34}|^2}{V\hbar \varepsilon_{0}}\frac{\delta_{34}-i\Gamma_5^{AV}/2}{\left(\delta_{3}-i\Gamma_4^{AV}/2\right)
\left(\delta_{34}-i\Gamma_5^{AV}/2\right)-|\Omega_{4}|^{2}}
\end{eqnarray}
\end{subequations}
and the following ones for the nonlinear susceptibilities,
\begin{widetext}
\begin{subequations}
\label{eq:suscgennonlinsym}
\begin{eqnarray}
&&\chi^{(3,sk)}_{P}=\frac{N |\mu_{32}|^4}{V\hbar^3 \varepsilon_{0}}
\frac{-\left(\delta_{12}-i\Gamma_1^{AV}/2\right)\left[\left|\delta_{12}-i\Gamma_1^{AV}/2\right|^2+|\Omega_{2}|^{2}\right]}{\left[\left(\delta_{1}-i\Gamma_2^{AV}/2\right)
\left(\delta_{12}-i\Gamma_1^{AV}/2\right)-|\Omega_{2}|^{2}\right]\left|\left(\delta_{1}-i\Gamma_2^{AV}/2\right)
\left(\delta_{12}-i\Gamma_1^{AV}/2\right)-|\Omega_{2}|^{2}\right|^2}, \\
&&\chi^{(3,sk)}_{T}=\frac{N |\mu_{34}|^4}{V\hbar^3 \varepsilon_{0}}
\frac{-\left(\delta_{34}-i\Gamma_5^{AV}/2\right)\left[\left|\delta_{34}-i\Gamma_5^{AV}/2\right|^2+|\Omega_{4}|^{2}\right]}{\left[\left(\delta_{3}-i\Gamma_4^{AV}/2\right)
\left(\delta_{34}-i\Gamma_5^{AV}/2\right)-|\Omega_{4}|^{2}\right]\left|\left(\delta_{3}-i\Gamma_4^{AV}/2\right)
\left(\delta_{34}-i\Gamma_5^{AV}/2\right)-|\Omega_{4}|^{2}\right|^2},\\
&&\chi^{(3,ck)}_{P}=\frac{N |\mu_{32}|^2|\mu_{34}|^2}{V\hbar^3 \varepsilon_{0}}
\frac{-\left(\delta_{12}-i\Gamma_1^{AV}/2\right)\left[\left|\delta_{34}-i\Gamma_5^{AV}/2\right|^2+|\Omega_{4}|^{2}\right]}{\left[\left(\delta_{1}-i\Gamma_2^{AV}/2\right)
\left(\delta_{12}-i\Gamma_1^{AV}/2\right)-|\Omega_{2}|^{2}\right]\left|\left(\delta_{3}-i\Gamma_4^{AV}/2\right)
\left(\delta_{34}-i\Gamma_5^{AV}/2\right)-|\Omega_{4}|^{2}\right|^2}, \\
&&\chi^{(3,ck)}_{T}=\frac{N |\mu_{32}|^2|\mu_{34}|^2}{V\hbar^3 \varepsilon_{0}}
\frac{-\left(\delta_{34}-i\Gamma_5^{AV}/2\right)\left[\left|\delta_{12}-i\Gamma_1^{AV}/2\right|^2+|\Omega_{2}|^{2}\right]}{\left[\left(\delta_{3}-i\Gamma_4^{AV}/2\right)
\left(\delta_{34}-i\Gamma_5^{AV}/2\right)-|\Omega_{4}|^{2}\right]\left|\left(\delta_{1}-i\Gamma_2^{AV}/2\right)
\left(\delta_{12}-i\Gamma_1^{AV}/2\right)-|\Omega_{2}|^{2}\right|^2}.
\end{eqnarray}
\end{subequations}
\end{widetext}
First of all we note that the expressions of the probe and trigger susceptibilities above are
completely symmetric. This means that probe and trigger experience the same linear and Kerr susceptibilities, as soon as the corresponding parameters
correspond, i.e., $\mu_{32}=\mu_{34}$,
$\Omega_1=\Omega_3$, $\Omega_2=\Omega_4$, $\delta_1=\delta_3$, $\delta_2=\delta_4$, $\Gamma_1^{AV}=\Gamma_5^{AV}$, $\Gamma_2^{AV}=\Gamma_4^{AV}$.
Moreover, the probe linear susceptibility of Eq.~(\ref{eq:probelinsym}a) and the self-Kerr susceptibility of Eq.~(\ref{eq:suscgennonlinsym}a)
coincide with the corresponding ones of the asymmetric case, Eq.~(\ref{eq:probelinasym}) and Eq.~(\ref{eq:suscgennonlin}a) respectively,
because the phenomenological decay rate $\Gamma_1^{AV}$ here plays just the same role of the phenomenological decay rate $\Gamma_3^{AV}$ of the
asymmetric scheme. This is not surprising, since the probe response in the absence of the trigger field is the same in the two \emph{M} scheme studied here.
Finally the cross-Kerr susceptibilities of the two schemes are generally different, both for the probe and the trigger, even though they possess
a similar structure. The main relevant difference between the two cross-Kerr susceptibilities is in the dependence of their real parts upon the detunings.
In fact, in the asymmetric case both real parts are proportional to the composite detuning $\delta_{14}=\delta_{12}+\delta_{34}$ (see Eqs.~(\ref{eq:delta})
and (\ref{eq:deltas})), so that one has a nonzero XPM as soon as one of the two $\Lambda$ subsystem is shifted from the two-photon resonance condition.
In the symmetric case instead, ${\rm Re}\{\chi^{(3,ck)}_{P}\}$ is proportional to $\delta_{12}$
and ${\rm Re}\{\chi^{(3,ck)}_{T}\}$ is proportional to $\delta_{34}$, and the two-photon resonance condition has to be
violated by \emph{both} $\Lambda$ subsystems if each field has to experience a nonzero XPM.

\subsection{Comparison with the Optical Bloch Equations} \label{sec:OBEsymmMain}

We now study the dynamics of the symmetric \emph{M} scheme of Fig.~\ref{fig:SMS} by means of the OBE,
which allow to describe spontaneous emission and dephasing more rigorously.
Due to the similarity of the symmetric and asymmetric \emph{M} schemes, we consider the same spontaneous emission and dephasing processes
described in subsection \ref{sec:OBEasymMain}. As a consequence, the master equation for the atomic density operator $\rho$ is again given by Eq.~(\ref{eq:blochgen}),
with the only difference that the Hamiltonian $H_{eff}^{AS}$ is replaced by the corresponding Hamiltonian $H_{eff}^{S}$
of the symmetric scheme, given by Eq.~(\ref{eq:hameffs}).
The corresponding system of OBE's for the mean values $\sigma_{ij}(t)\equiv \langle \hat{\sigma}_{ij}(t)\rangle \equiv
\rho_{ji}(t)$ is displayed in Appendix~\ref{app:OBEsymm} as Eqs.~(\ref{eq:blochpopSymm}) and (\ref{eq:blochcohSymm}), where we have used the definitions of
Eqs.~(\ref{eq:decay1})-(\ref{eq:deph}).

Also in this symmetric case, the OBE are less suited for an approximate analytical treatment with respect to the AV equations
of the preceding subsection. In fact, if we consider the conditions $|\Omega_1/\Omega_2| \ll 1$ and $|\Omega_3/\Omega_4| \ll 1$
and, consistently with Eq.~(\ref{eq:popassumption}), we assume that
\begin{subequations}
\label{eq:approxsym}
\begin{eqnarray}
\sigma_{33} &\approx& 1, \\
\sigma_{jj} &\approx& 0, \;\;\;\;  j=1,2,4,5,
\end{eqnarray}
\end{subequations}
at the steady state, it is possible to see that by inserting Eqs.~(\ref{eq:approx}) into Eqs.~(\ref{eq:blochcohSymm}) for the
coherences, one gets a satisfactory expression for the probe linear susceptibility only. To be more specific, only the approximate linear susceptibility
fits well with the numerical solution of the OBE. It is not easy to derive analytical expressions from
Eqs.~(\ref{eq:blochpopSymm}) and (\ref{eq:blochcohSymm}) for the nonlinear susceptibilities which would be as simple as those
of Eqs.~(\ref{eq:suscgennonlinsym}) and which would reproduce in the same way the exact numerical solution of the OBE within the EIT regime.
Again, one could exactly solve analytically the OBE, but the resulting expressions are very cumbersome and not physically transparent as those
of Eqs.~(\ref{eq:suscgennonlinsym}). For this reason we will analytically derive from the OBE the probe linear susceptibility only, and we will
then use the OBE only for the numerical determination of the atomic steady state. In addition, deriving this result will enable us to draw a formal
analogy between the AV and OBE treatments (see Eqs.~(\ref{eq:reinter2}) below).

The probe and trigger susceptibilities are now defined as
\begin{subequations}
\label{eq:avchi3}
\begin{eqnarray}
&& \chi_{P} = \frac{N \mu_{32}}{V\varepsilon_{0}{\mathcal E}_1}\sigma_{32}=-\frac{N |\mu_{32}|^2}{V\hbar \varepsilon_{0}\Omega_1}
\sigma_{32}, \\
&& \chi_{T} = \frac{N \mu_{34}}{V\varepsilon_{0}{\mathcal E}_3}\sigma_{34}=-\frac{N |\mu_{34}|^2}{V\hbar \varepsilon_{0}\Omega_3}
\sigma_{34}.
\end{eqnarray}
\end{subequations}
Using Eqs.~(\ref{eq:approxsym}) and performing a series expansion at the lowest order in the probe and trigger fields,
we arrive at the following expressions for the probe and trigger linear susceptibilities
\begin{subequations}
\label{eq:suscsym}
\begin{eqnarray}\label{eq:suscprobesym}
&& \chi^{(1)}_{P}=\frac{N |\mu_{32}|^2}{V\hbar \varepsilon_{0}}\frac{\delta_{12}-i\gamma_{13}/2}{\left[\delta_{12}-i\gamma_{13}/2\right]
\left[\delta_{1}-i\left(\Gamma_{2}+\gamma_{12}\right)/2\right]-|\Omega_{2}|^{2}}, \\
&& \chi^{(1)}_{T}=\frac{N |\mu_{34}|^2}{V\hbar \varepsilon_{0}}\frac{\delta_{34}-i\gamma_{53}/2}{\left[\delta_{34}-i\gamma_{53}/2\right]
\left[\delta_{3}-i\left(\Gamma_{4}+\gamma_{54}\right)/2\right]-|\Omega_{4}|^{2}}.
\end{eqnarray}
\end{subequations}
The probe linear susceptibility of Eq.~(\ref{eq:suscprobesym}) coincides with that of the asymmetric \emph{M} scheme of Eq.~(\ref{eq:suscprobe}),
as it must be, since the
linear properties of the probe in the two \emph{M} schemes are identical.
Moreover, as noted before, due to the symmetry of the scheme, this probe linear susceptibility coincides with that of the
trigger of Eq.~(\ref{eq:suscsym}b) when the corresponding parameters coincide, i.e., $\mu_{32}=\mu_{34}$,
$\delta_1=\delta_3$, $\delta_2=\delta_4$, $\Omega_2=\Omega_4$,
$\Gamma_2=\Gamma_4$, $\gamma_{54}=\gamma_{12}$ and $\gamma_{53}=\gamma_{13}$.
By comparing Eqs.~(\ref{eq:suscsym}) with Eqs.~(\ref{eq:probelinsym}), one can also see that
the AV and OBE predictions for the linear susceptibilities again coincide provided that
the phenomenological decay rates $\Gamma_i^{AV}$ are appropriately interpreted, i.e.,
\begin{subequations}
\label{eq:reinter2}
\begin{eqnarray}
\Gamma_2^{AV}& \leftrightarrow &\Gamma_{2}+\gamma_{12}, \\
\Gamma_1^{AV}& \leftrightarrow &\gamma_{13},\\
\Gamma_4^{AV}& \leftrightarrow &\Gamma_{4}+\gamma_{54}, \\
\Gamma_5^{AV}& \leftrightarrow &\gamma_{53}.
\end{eqnarray}
\end{subequations}
This shows again that the intuitive interpretation of the phenomenological decay rates $\Gamma_i^{AV}$
as spontaneous emission total decay rates for the
excited states, and as dephasing rates in the case of ground state sublevels, is essentially correct.

\begin{figure}[t]
\begin{center}
\includegraphics[scale=0.45]{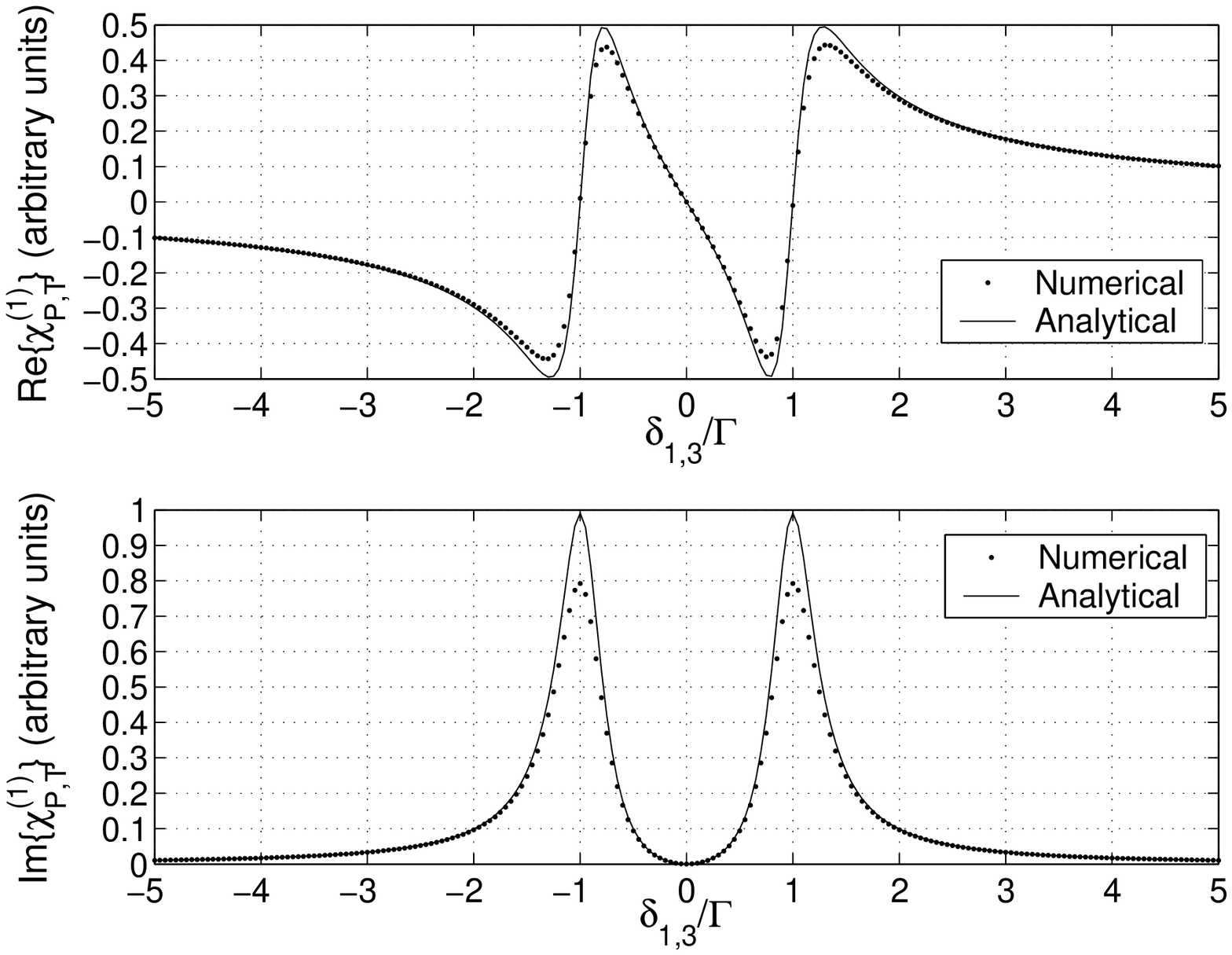}
\caption{Comparison of the numerical solution (dotted line) of the OBE with the analytical prediction
of Eqs.~(\protect\ref{eq:probelinsym}) (full line) for the real part (above) and imaginary part (below) of both probe and trigger linear
susceptibilities versus their respective normalized probe detunings $\delta_{1}/\Gamma$ and $\delta_{3}/\Gamma$. Probe and trigger susceptibilities
exactly overlap because we consider the perfectly symmetric situation $\Gamma_{2}^{AV}=\Gamma_{2}=\Gamma_{4}^{AV}=\Gamma_{4} = \Gamma =  2\pi\times 6$ MHz,
$\Omega_{1}=\Omega_{3}=0.08\Gamma$, $\Omega_{2} = \Omega_{4} = \Gamma$, $\delta_2=\delta_4=0$, $\forall$ i,j
$\gamma_{ij}=\Gamma_{1}^{AV}=\Gamma_{5}^{AV}=10^{-4}\Gamma$, $\mu_{32}=\mu_{34}$, which guarantees perfect group velocity matching.}
\label{fig:8}
\end{center}
\end{figure}

\begin{figure}[t]
\begin{center}
\includegraphics[scale=0.45]{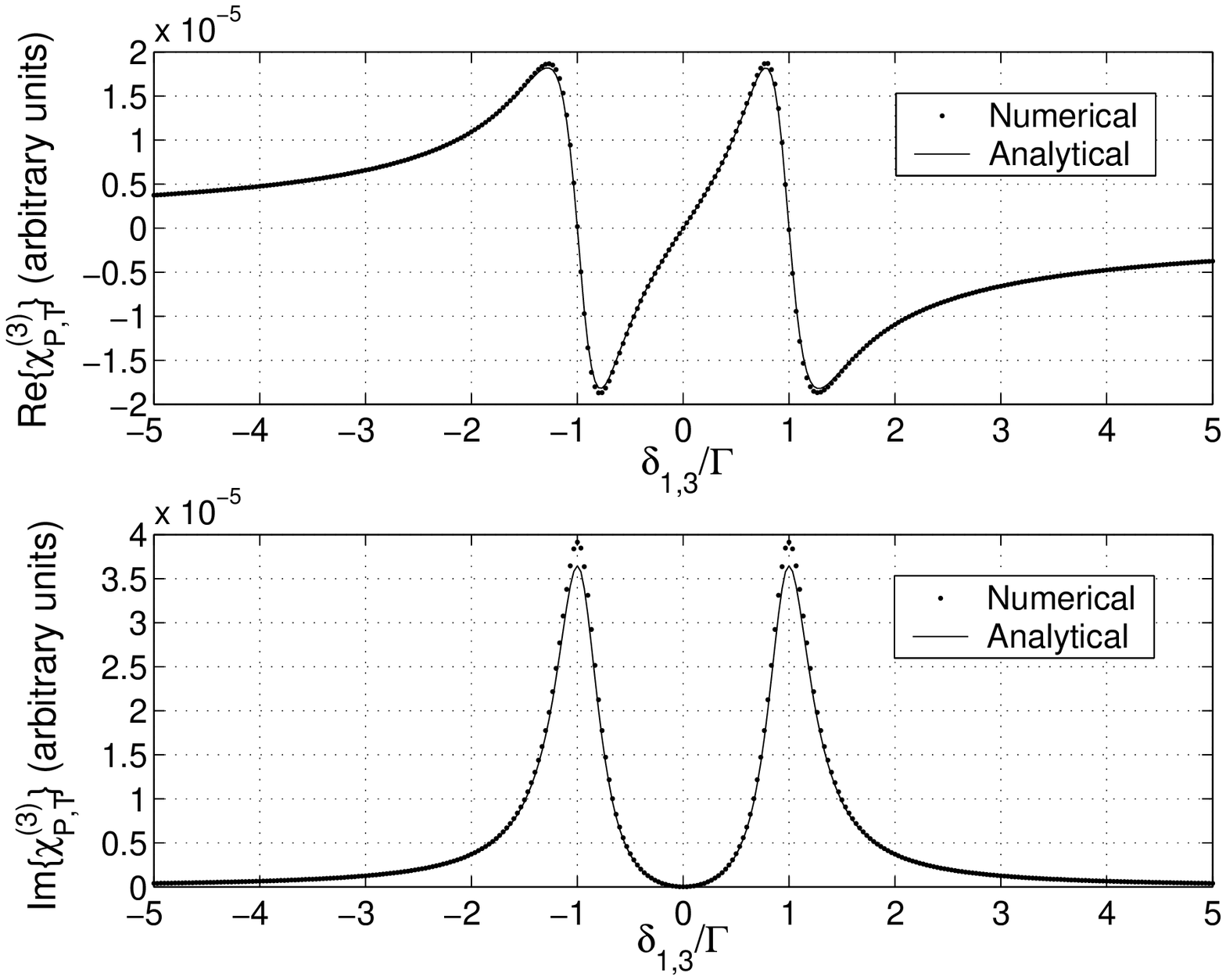}
\caption{Comparison of the numerical solution (dotted line) of the OBE with the analytical prediction
of Eqs.~(\protect\ref{eq:suscgennonlinsym}c,d) (full line) for the real part (above) and imaginary part (below) of both probe and trigger cross-Kerr
susceptibilities versus their respective normalized probe detunings $\delta_{1}/\Gamma$ and $\delta_{3}/\Gamma$. Probe and trigger susceptibilities
exactly overlap because we consider a perfectly symmetric situation: $\Gamma_{2}^{AV}=\Gamma_{2}=\Gamma_{4}^{AV}=\Gamma_{4} = \Gamma =  2\pi\times 6$ MHz,
$\Omega_{2} = \Omega_{4} = \Gamma$, $\delta_2=\delta_4=0$, $\forall$ i,j
$\gamma_{ij}=\Gamma_{1}^{AV}=\Gamma_{5}^{AV}=10^{-4}\Gamma$; moreover we have chosen $\Omega_{1}=0.002\Gamma$, $\Omega_{3}=0.08\Gamma$ in the case of the
$\chi^{(3,ck)}_{P}$ plot, and \emph{viceversa} $\Omega_{3}=0.002\Gamma$, $\Omega_{1}=0.08\Gamma$ in the case of the $\chi^{(3,ck)}_{T}$ plot,
in order to reduce as much as possible the influence of the self-Kerr susceptibilities.}
\label{fig:9}
\end{center}
\end{figure}

\begin{figure}[t]
\begin{center}
\includegraphics[scale=0.45]{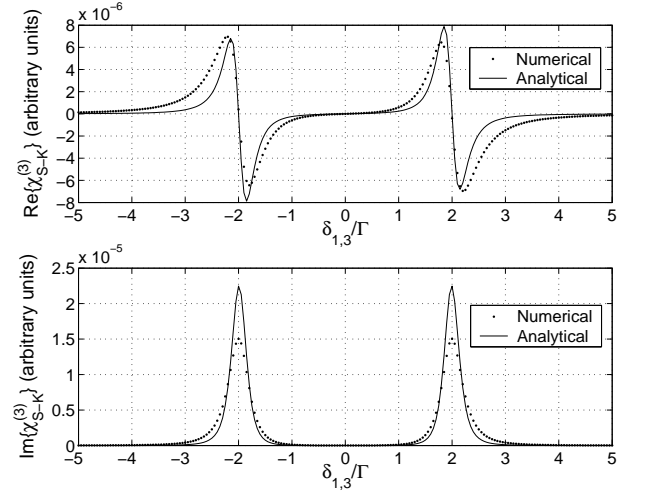}
\caption{Comparison of the numerical solution (dotted line) of the OBE with the analytical prediction
of Eqs.~(\protect\ref{eq:suscgennonlinsym}a,b) (full line) for the real part (above) and imaginary part (below) of both probe and trigger self-Kerr
susceptibilities versus their respective normalized probe detunings $\delta_{1}/\Gamma$ and $\delta_{3}/\Gamma$. Probe and trigger susceptibilities
exactly overlap because we consider a perfectly symmetric situation: $\Gamma_{2}^{AV}=\Gamma_{2}=\Gamma_{4}^{AV}=\Gamma_{4} = \Gamma =  2\pi\times 6$ MHz,
$\Omega_{2} = \Omega_{4} = 2\Gamma$, $\delta_2=\delta_4=0$, $\forall$ i,j
$\gamma_{ij}=\Gamma_{1}^{AV}=\Gamma_{5}^{AV}=10^{-4}\Gamma$; moreover we have chosen $\Omega_{1}=0.4\Gamma$, $\Omega_{3}=0.004\Gamma$ in the case of the
$\chi^{(3,sk)}_{P}$ plot, and \emph{viceversa} $\Omega_{3}=0.4\Gamma$, $\Omega_{1}=0.004\Gamma$ in the case of the $\chi^{(3,sk)}_{T}$ plot,
in order to reduce as much as possible the influence of the cross-Kerr susceptibilities.}
\label{fig:9bis}
\end{center}
\end{figure}

We then consider the numerical solution of the OBE and we compare it with the analytical treatment based on the
AV approach. The numerical calculations are again performed in the limits discussed
above, i.e., $|\Omega_1|,|\Omega_2| \ll |\Omega_3|, |\Omega_4|$ and we stay near the Raman resonance for both the probe and the trigger.
In Figs.~\ref{fig:8}-\ref{fig:9bis} we compare the analytical solutions of Eqs.~(\ref{eq:probelinsym}) and Eqs.~(\ref{eq:suscgennonlinsym}) with the
numerical solution of the complete set of Bloch equations given in the Appendix B. Fig.~\ref{fig:8} shows the linear susceptibilities
and refers to a perfectly symmetric situation
between probe and trigger, i.e., $\Gamma_{2}^{AV}=\Gamma_{2}=\Gamma_{4}^{AV}=\Gamma_{4} = \Gamma =  2\pi\times 6$ MHz,
$\Omega_{1}=\Omega_{3}=0.08\Gamma$, $\Omega_{2} = \Omega_{4} = \Gamma$, $\delta_2=\delta_4=0$, $\forall$ i,j
$\gamma_{ij}=\Gamma_{1}^{AV}=\Gamma_{5}^{AV}=10^{-4}\Gamma$. As a consequence, the probe and trigger linear susceptibilities as a function of the
respective detunings $\delta_1$ and $\delta_3$ are two indistinguishable curves. In such a case,
group velocity matching is automatically guaranteed whenever $\mu_{32}=\mu_{34}$.
Fig.~\ref{fig:8} shows that Eqs.~(\ref{eq:probelinsym}) work very well, except when
the detunings correspond to the maximum probe (or trigger) absorption. In such a case, the detunings match the Rabi frequencies of the
two pumps, and the probe (or trigger) field is in resonance with a single atomic transition. The atoms are significantly pumped to the excited levels
and the population assumption of Eq.~(\ref{eq:popassumption}) is no more fulfilled.

Fig.~\ref{fig:9} shows the cross-Kerr susceptibilities again in a perfectly symmetric situation between probe and trigger
so that their plots as a function of the respective detunings $\delta_1$ and $\delta_3$ exactly coincide. However,
in order to reduce as much as possible the influence due to the simultaneous presence of the self-Kerr susceptibility,
we have considered a probe Rabi frequency $\Omega_{1}$ much smaller than that of the trigger in the $\chi^{(3,ck)}_{P}$ plot and
\emph{viceversa} a trigger Rabi frequency $\Omega_{3}$ much smaller than that of the probe in the $\chi^{(3,ck)}_{T}$ plot. To be more precise,
Fig.~\ref{fig:9} refers to $\Gamma_{2}^{AV}=\Gamma_{2}=\Gamma_{4}^{AV}=\Gamma_{4} = \Gamma =  2\pi\times 6$ MHz,
$\Omega_{2} = \Omega_{4} = \Gamma$, $\delta_2=\delta_4=0$, $\forall$ i,j
$\gamma_{ij}=\Gamma_{1}^{AV}=\Gamma_{5}^{AV}=10^{-4}\Gamma$, and $\Omega_{1}=0.002\Gamma$, $\Omega_{3}=0.08\Gamma$ in the case of the
$\chi^{(3,ck)}_{P}$ plot, and to $\Omega_{3}=0.002\Gamma$, $\Omega_{1}=0.08\Gamma$ in the case of the $\chi^{(3,ck)}_{T}$ plot.
We can see from Fig.~\ref{fig:9} that the AV approach gives a satisfactory description also for the cross-Kerr susceptibility, except when
probe or trigger absorption is maximum, as it happens for the linear case.

Finally Fig.~\ref{fig:9bis} shows the self-Kerr susceptibilities again in a perfectly symmetric situation between probe and trigger.
As a consequence their plots versus the respective detunings $\delta_1$ and $\delta_3$ exactly coincide. Here,
in order to reduce the influence due to the simultaneous presence of the cross-Kerr susceptibility,
we have considered a trigger Rabi frequency $\Omega_{3}$ much smaller than that of the probe in the $\chi^{(3,sk)}_{P}$ plot and
\emph{viceversa} a probe Rabi frequency $\Omega_{1}$ much smaller than that of the trigger in the $\chi^{(3,sk)}_{T}$ plot. The parameters are the same as
in Fig.~\ref{fig:9}, except that here we have chosen $\Omega_{2} = \Omega_{4} = 2\Gamma$,
$\Omega_{1}=0.4\Gamma$, $\Omega_{3}=0.004\Gamma$ in the case of the
$\chi^{(3,sk)}_{P}$ plot, and $\Omega_{3}=0.4\Gamma$, $\Omega_{1}=0.004\Gamma$ in the case of the $\chi^{(3,sk)}_{T}$ plot.
The agreement between the AV prediction and the numerical solution of the OBE is satisfactory.

\subsection{Group velocity matching}

\begin{figure}[t]
\begin{center}
\includegraphics[scale=0.45]{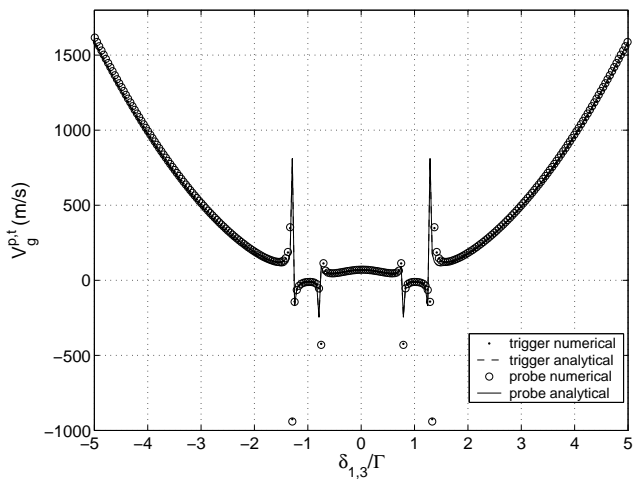}
\caption{Group velocity of the probe and trigger pulses versus the normalized detunings $\delta_1/\Gamma=\delta_3/\Gamma$.
Lines denote the analytical predictions of Eqs.~(\protect\ref{eq:vgdef}) and (\protect\ref{eq:vgsym})
(the full line refers to the probe and the dashed line to the trigger).
Circles and dots refer to the numerical solution of the OBE for the probe and trigger group velocity, respectively.
Parameters correspond to the perfectly symmetric situation considered in Fig.~\ref{fig:8}, that is,
$\Gamma_{2}^{AV}=\Gamma_{2}=\Gamma_{4}^{AV}=\Gamma_{4} = \Gamma =  2\pi\times 6$ MHz,
$\Omega_{1}=\Omega_{3}=0.08\Gamma$, $\Omega_{2} = \Omega_{4} = \Gamma$, $\delta_2=\delta_4=0$, $\forall$ i,j
$\gamma_{ij}=\Gamma_{1}^{AV}=\Gamma_{5}^{AV}=10^{-4}\Gamma$, and we have chosen $\mu_{32}=\mu_{34}= 10^{-29}$ C $\cdot$m, and
$N/V=3.0 \times 10^{13}$ cm$^{-3}$. Due to symmetry, one has perfect group velocity matching within a large interval of values for the detunings.}
\label{fig:10}
\end{center}
\end{figure}

As we have seen in subsection \ref{sec:vgmatch}, the condition of group velocity matching is of fundamental importance
for achieving a large cross-phase modulation between probe and trigger fields.
It is evident from the inherent symmetry of the present scheme that the condition of equal probe and trigger group velocities
is automatically achieved when the corresponding parameters are equal i.e., $\mu_{32}=\mu_{34}$, $\delta_1=\delta_3$, $\delta_2=\delta_4$, $\Omega_2=\Omega_4$,
$\Gamma_2=\Gamma_4$, $\gamma_{54}=\gamma_{12}$ and $\gamma_{53}=\gamma_{13}$, $\omega_1 \simeq \omega_3$.
This is the main advantage of the symmetric \emph{M} scheme over the asymmetric one. As we have seen above, the group velocity
of a pulse is given by
\begin{equation} \label{eq:vgdef}
v_g^i = \frac{c}{1+n_g^i}, \; \; i=P,T,
\end{equation}
where the group index $n_g^i$ is given by Eq.~(\ref{eq:ng}). The contribution of the nonlinear susceptibilities to $v_g$
is negligible with respect to that of the linear one, which is nonzero for both probe and trigger in this case (see Eqs.~(\ref{eq:suscgensym})).
Therefore, approximating $\chi$ with the linear contribution $\chi^{(1)}$ and inserting Eqs.~(\ref{eq:probelinsym}) into the definition (\ref{eq:ng}),
one gets the following expressions for the two group indices
\begin{subequations}
\label{eq:vgsym}
\begin{eqnarray}
&& n_g^P = \frac{N |\mu_{32}|^2}{2V\hbar \varepsilon_{0}}{\rm Re}\left\{\frac{d_1}{d_1 d_2-|\Omega_{2}|^{2}}+
 \frac{\omega_1\left(d_1^2+|\Omega_{2}|^{2}\right)}{\left(d_1 d_2-|\Omega_{2}|^{2}\right)^2}\right\}, \\
&& n_g^T = \frac{N |\mu_{34}|^2}{2V\hbar \varepsilon_{0}}{\rm Re}\left\{\frac{d_5}{d_5 d_4-|\Omega_{4}|^{2}}+
 \frac{\omega_3\left(d_5^2+|\Omega_{4}|^{2}\right)}{\left(d_5 d_4-|\Omega_{4}|^{2}\right)^2}\right\},
\end{eqnarray}
\end{subequations}
where we have used the definitions of Eqs.~(\ref{eq:complexdet}) for $d_j$, $j=1,2,4,5$. The symmetry between probe and trigger discussed above is evident
also in these expressions. Eqs.~(\ref{eq:vgdef}) and (\ref{eq:vgsym}) are now compared with the corresponding
ones obtained from the integration of the full set of Bloch equations of
Appendix~\ref{app:OBEsymm}. The comparison is shown in Fig.~\ref{fig:10}, which refers to the completely symmetric situation between probe and trigger
defined above and therefore shows exact group velocity matching for all values of the detunings $\delta_{1}=\delta_{3}$.
Fig.~\ref{fig:10} shows an excellent agreement between analytical and numerical results. The only points in
which the two curves do not coincide exactly are when the detunings match the Rabi frequencies of the two pumping field. In fact in this
conditions the fields are in resonance with a single atomic transition and the atoms are pumped to the excited levels. The other points
that determine the disagreement are in the vicinity of the peaks. In fact in these regions the derivatives are small, because of the change
in slope of the real part of the susceptibilities. Hence, the group index of Eqs.~(\ref{eq:vgsym}) is small,
and the group velocity jumps near $c$.

\begin{figure}[htb]
\begin{center}
\includegraphics[scale=0.45]{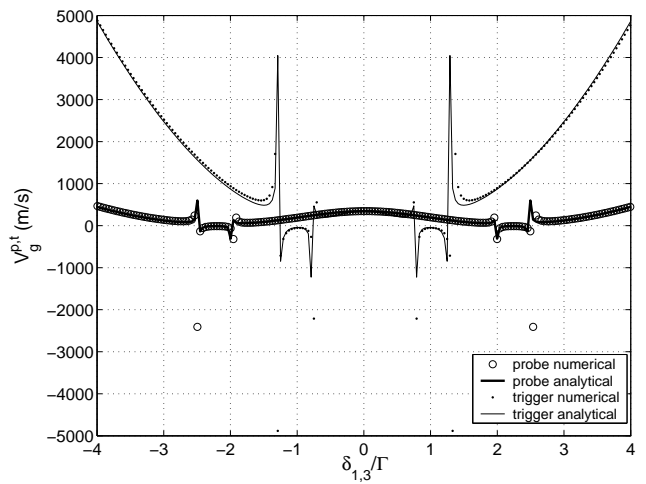}
\caption{Group velocity of the probe and trigger pulses versus the normalized detunings $\delta_1/\Gamma=\delta_3/\Gamma$.
Full lines denote the analytical predictions of Eqs.~(\protect\ref{eq:vgdef}) and (\protect\ref{eq:vgsym})
(the thick line refers to the probe and the thin line to the trigger).
Circles and dots refer to the numerical solution of the OBE for the probe and trigger group velocity, respectively.
Parameters correspond to the five--level scheme derived from the $D_{1}$ and $D_{2}$ lines of the $^{87}$Rb spectrum described in the text,
$\Gamma_2^{AV}=\Gamma_{2} \simeq 36$ MHz, $\Gamma_4^{AV}=\Gamma_{4}\simeq 38$ MHz,
$\mu_{32}=1.27 \times 10^{-29}$ C$\cdot$m, $\mu_{34}=5.7 \times 10^{-30}$ C$\cdot$m,
$\Omega_{4} = \Gamma$, $\Omega_{2} = 2.22\Gamma$, $\Omega_{1}=\Omega_{3}=0.08\Gamma$, $\delta_2=\delta_4=0$, $\forall$ i,j
$\gamma_{ij}=\Gamma_{1}^{AV}=\Gamma_{5}^{AV}=10^{-4}\Gamma$, $N/V=3.0\times 10^{13}$ cm$^{-3}$. The asymmetry between the two dipole moment
matrix elements has been compensated by adjusting the value of $\Omega_{2}$. In this way group velocity matching is achieved
within the entire EIT window. }
\label{fig:11}
\end{center}
\end{figure}

In some cases, the perfectly symmetric conditions guaranteeing group velocity matching, i.e.,
$\mu_{32}=\mu_{34}$, $\delta_1=\delta_3$, $\delta_2=\delta_4$, $\Omega_2=\Omega_4$,
$\Gamma_2=\Gamma_4$, $\gamma_{54}=\gamma_{12}$ and $\gamma_{53}=\gamma_{13}$, are difficult to realize in practice. In fact,
sometimes it may be convenient to use transitions with different Clebsch-Gordan coefficients, yielding therefore a
significant discrepancy between $\mu_{32}$ and $\mu_{34}$. The other symmetry conditions above are less problematic because detunings and Rabi frequencies can always be
made equal by the experimenter, and moreover decay and dephasing rates, even though not perfectly equal, are often comparable to each other.
Just to give an example, one could implement the symmetric \emph{M} scheme of Fig.~\ref{fig:SMS} by using the $D_{1}$ and $D_{2}$ line of the $^{87}$Rb spectrum.
The Zeeman sublevels $|5P_{1/2}F=1,m=0\rangle$ and $|5P_{3/2}F=1,m=0\rangle$
could be chosen as levels $|2\rangle$ and $|4\rangle$, respectively, while the Zeeman sublevels $|5S_{1/2}F=1,m=-1\rangle$, $|5S_{1/2}F=2,m=1\rangle$
and $|5P_{1/2}F=1,m=1\rangle$ could chosen as levels $|1\rangle$, $|3\rangle$ and $|5\rangle$, respectively (see also Ref.~\cite{nostro} for a similar choice).
For these levels the atomic transitions related to the probe and trigger
have dipole moment matrix elements $\mu_{32}$, $\mu_{34}$ differing by a factor $\sqrt{10}$, violating therefore the symmetry condition.
It is evident however that this slight asymmetry can be compensated (so that group velocity matching can be still achieved in a restricted
but still useful range of detunings) by properly adjusting the Rabi frequencies of the tuner field $\Omega_{4}$ and of the coupling field $\Omega_{2}$,
which will be no more equal. In fact, by imposing group velocity matching at the center of the transparency
window, i.e., for $\delta_{12}=\delta_{34}=0$, we derive the condition
\begin{equation}\label{eq:corr}
\Omega_{2} = \alpha\Omega_{4},
\end{equation}
where the \emph{correction factor} $\alpha$ is given by
\begin{equation}\label{eq:alpha}
\alpha=\sqrt{\frac{|\mu_{32}|^{2}}{|\mu_{34}|^{2}}\frac{\omega_{1}}{\omega_{3}}}.
\end{equation}
As shown in Fig.~\ref{fig:11}, if the adjustment condition of Eqs.~(\ref{eq:corr}) and (\ref{eq:alpha}) is taken into account,
one still gets equal probe and trigger group velocities in the case of the $^{87}$Rb five--level scheme specified above,
at least within the entire EIT window.

Finally, we note also that the problem with the pulse \textit{propagation}, present in the asymmetric arrangement (c.f. Sec.~\ref{sec:pulseprop}) is absent in the symmetric arrangement. This is because the dispersion, and the related group velocity reduction have its origin in the \textit{linear} part of the susceptibility for both, probe and trigger pulses. Hence there is no apparent singularity in the trigger group velocity, which in asymmetric case stemmed from the nonlinear origin of trigger dispersion.

\section{Conclusions}\label{sec:conc}

We have studied a five-level atomic system in two different but related $M$-configurations. We focused on the nonlinear properties of the system
and specifically on the conditions for the optimization of the cross-phase modulation between two weak fields of interest, which we have named probe and trigger
fields. Both systems have been studied from a
semiclassical point of view, i.e., by describing all the fields in terms of their Rabi frequencies. We have
seen that both linear and nonlinear properties are well described by an approach based on amplitude variables,
which has been shown to reproduce well the numerical solution of the exact optical Bloch equations describing the system.

Both the asymmetric and the symmetric \emph{M} scheme are able to provide a giant cross-Kerr modulation, which may be useful for many
applications. Both \emph{M} schemes can be seen as a ``duplication'' of the usual three-level $\Lambda$ scheme at the basis of EIT, one for the probe
and one for the trigger fields. In the asymmetric scheme, only the probe drives a significantly populated transition and a large cross-Kerr effect is obtained
when either the probe or the trigger is slightly detuned from the two-photon resonance condition.
The corresponding nonlinear phase shift, yielding for example the conditional phase shift of Eq.~(\ref{eq:def_cps})
of a quantum phase gate for photonic qubits, can become very large, especially
when the probe and trigger group velocities, slowed down by EIT, become equal. In the asymmetric scheme, this group velocity matching can be achieved
by properly adjusting the detuning and the intensity of the control field of the trigger $\Lambda$ system.
In the symmetric \emph{M} scheme, the atomic population is equally shared by the probe and trigger transitions.
Adjusting the corresponding parameters (Rabi frequencies, detunings) so that the two $\Lambda$ systems become identical,
probe and trigger experience the same interaction with the atomic medium
and group velocity matching is achieved automatically. In this case a significant nonlinear cross-phase modulation is achieved
only if both $\Lambda$ schemes are slightly and equally detuned from two-photon resonance, so to remain still within the transparency window.
In fact, due to EIT, the susceptibility vanishes at all orders at the exact two-photon resonance condition.

\begin{widetext}

\appendix

\section{Optical Bloch Equations - Asymmetric Case} \label{app:OBEasym}

From Eqs.~(\ref{eq:hameff}) and (\ref{eq:blochgen}), and using the definitions of Eqs.~(\ref{eq:decay1})-(\ref{eq:deph}), one gets the following set of
equations for the atomic populations $\sigma_{ii}$

\begin{subequations}
\label{eq:blochpopAsym}
\begin{eqnarray}
\dot{\sigma}_{11} &=& i\Omega_{1}\sigma_{21} -i\Omega^{\star}_{1}\sigma_{12} +\Gamma_{41}\sigma_{44} + \Gamma_{21}\sigma_{22},\\
\dot{\sigma}_{22} &=& -i\Omega_{1}\sigma_{21} +i\Omega^{\star}_{1}\sigma_{12} -i\Omega_{2}\sigma_{23} +i\Omega^{\star}_{2}\sigma_{32} -\Gamma_{2}\sigma_{22},\\
\dot{\sigma}_{33} &=& i\Omega_{3}\sigma_{43} -i\Omega^{\star}_{3}\sigma_{34} +i\Omega_{2}\sigma_{23} -i\Omega^{\star}_{2}\sigma_{32} +\Gamma_{43}\sigma_{44}
+\Gamma_{23}\sigma_{22},\\
\dot{\sigma}_{44} &=& i\Omega^{\star}_{3}\sigma_{34} -i\Omega_{3}\sigma_{43} -i\Omega_{4}\sigma_{45} +i\Omega^{\star}_{4}\sigma_{54} -\Gamma_{4}\sigma_{44},\\
\dot{\sigma}_{55} &=& i\Omega_{4}\sigma_{45} -i\Omega^{\star}_{4}\sigma_{54} +\Gamma_{25}\sigma_{22} +\Gamma_{45}\sigma_{44},
\end{eqnarray}
\end{subequations}
and the following set of equations for the atomic coherences $\sigma_{ij}$, $i\neq j$,
\begin{subequations}
\label{eq:blochcohAsym}
\begin{eqnarray}
\dot{\sigma}_{12} &=& -i\delta_{1}\sigma_{12} +i\Omega_{1}(\sigma_{22} -\sigma_{11}) -i\Omega_{2}\sigma_{13} -\frac{\Gamma_{2} +\gamma_{12}}{2}\sigma_{12},\\
\dot{\sigma}_{13} &=& -i\delta_{12}\sigma_{13} +i\Omega_{1}\sigma_{23} -i\Omega^{\star}_{3}\sigma_{14} -i\Omega^{\star}_{2}\sigma_{12} -\frac{\gamma_{13}}
{2}\sigma_{13}, \\
\dot{\sigma}_{14} &=& -i\delta_{13}\sigma_{14} +i\Omega_{1}\sigma_{24} -i\Omega_{3}\sigma_{13} -i\Omega_{4}\sigma_{15} -\frac{\gamma_{14}+\Gamma_{4}}{2}
\sigma_{14}, \\
\dot{\sigma}_{15} &=& -i\delta_{14}\sigma_{15} +i\Omega_{1}\sigma_{25} -i\Omega^{\star}_{4}\sigma_{14} -\frac{\gamma_{15}}{2}\sigma_{15}, \\
\dot{\sigma}_{23} &=& i\delta_{2}\sigma_{23} +i\Omega^{\star}_{1}\sigma_{13} -i\Omega^{\star}_{3}\sigma_{24} +i\Omega^{\star}_{2}(\sigma_{33}-\sigma_{22})
-\frac{\Gamma_{2}+\gamma_{23}}{2}\sigma_{23}, \\
\dot{\sigma}_{24} &=& i\delta_{23}\sigma_{24} -i\Omega_{3}\sigma_{23} -i\Omega_{4}\sigma_{25} +i\Omega^{\star}_{2}\sigma_{34} +i\Omega^{\star}_{1}\sigma_{14}
-\frac{\Gamma_{2}+\Gamma_{4}+\gamma_{24}}{2}\sigma_{24}, \\
\dot{\sigma}_{25} &=& i\delta_{24}\sigma_{25} +i\Omega^{\star}_{1}\sigma_{15} +i\Omega^{\star}_{2}\sigma_{35} -i\Omega^{\star}_{4}\sigma_{24}
-\frac{\Gamma_{2}+\gamma_{25}}{2}\sigma_{25}, \\
\dot{\sigma}_{34} &=& -i\delta_{3}\sigma_{34} +i\Omega_{3}(\sigma_{44}-\sigma_{33}) +i\Omega_{2}\sigma_{24} -i\Omega_{4}\sigma_{35} -\frac{\Gamma_{4}
+\gamma_{34}}{2}\sigma_{23}, \\
\dot{\sigma}_{35} &=& -i\delta_{34}\sigma_{35} +i\Omega_{3}\sigma_{45} +i\Omega_{2}\sigma_{25} -i\Omega^{\star}_{4}\sigma_{34}
-\frac{\gamma_{35}}{2}\sigma_{35}, \\
\dot{\sigma}_{45} &=& i\delta_{4}\sigma_{45} +i\Omega^{\star}_{3}\sigma_{35} +i\Omega^{\star}_{4}(\sigma_{55}-\sigma_{44}) -\frac{\Gamma_{4}+\gamma_{45}}{2}\sigma_{45},
\end{eqnarray}
\end{subequations}
where we have also defined the composite detunings $\delta_{23}=\delta_2-\delta_3$, $\delta_{24}=\delta_2-\delta_3+\delta_4$, and $\delta_{34}=\delta_3-\delta_4$.

\section{Optical Bloch Equations - Symmetric Case} \label{app:OBEsymm}

From Eqs.~(\ref{eq:hameffs}) and (\ref{eq:blochgen}), and using the definitions of Eqs.~(\ref{eq:decay1})-(\ref{eq:deph}), one gets the following set of
equations for the atomic populations $\sigma_{ii}$
\begin{subequations}
\label{eq:blochpopSymm}
\begin{eqnarray}
\dot{\sigma}_{11} &=& \imath\Omega_{2}\sigma_{21}-\imath\Omega^{\star}_{2}\sigma_{12} + \Gamma_{41}\sigma_{44} + \Gamma_{21}\sigma_{22},\\
\dot{\sigma}_{22} &=& \imath\Omega_{2}\sigma_{12} - \imath\Omega^{\star}_{2}\sigma_{21} - \imath\Omega_{1}\sigma_{23} + \imath\Omega^{\star}_{1}
\sigma_{32}  -\Gamma_{2}\sigma_{22},\\
\dot{\sigma}_{33} &=& \imath\Omega_{3}\sigma_{43} - \imath\Omega^{\star}_{3}\sigma_{34} + \imath\Omega_{1}\sigma_{23} -
\imath\Omega^{\star}_{1}\sigma_{32} + \Gamma_{43}\sigma_{44} + \Gamma_{23}\sigma_{2},\\
\dot{\sigma}_{44} &=& \imath\Omega^{\star}_{3}\sigma_{34} - \imath\Omega_{3}\sigma_{43} - \imath\Omega_{4}\sigma_{45} + \imath\Omega^{\star}_{4}\sigma_{54}
-\Gamma_{4}\sigma_{44},\\
\dot{\sigma}_{55} &=& \imath\Omega_{4}\sigma_{45}-\imath\Omega^{\star}_{4}\sigma_{54} + \Gamma_{45}\sigma_{44} + \Gamma_{25}\sigma_{22},
\end{eqnarray}
\end{subequations}
while the equations for the coherences are
\begin{subequations}
\label{eq:blochcohSymm}
\begin{eqnarray}
\dot{\sigma}_{12} &=& -\imath\delta_{2}\sigma_{12}+\imath\Omega_{2}(\sigma_{22}-\sigma_{11})-
\imath\Omega_{1}\sigma_{13} -\frac{\Gamma_{2}+\gamma_{12}}{2}\sigma_{12},\\
\dot{\sigma}_{13} &=& \imath (\delta_{1}-\delta_{2})\sigma_{13} +\imath\Omega_{2}\sigma_{23} -\imath\Omega^{\star}_{3}\sigma_{14}
-\imath\Omega^{\star}_{1}\sigma_{12} -\frac{\gamma_{13}}{2}\sigma_{13},\\
\dot{\sigma}_{14} &=& \imath(\delta_{1}-\delta_{2}-\delta_{3})\sigma_{14} +\imath\Omega^{\star}_{2}\sigma_{24} -\imath\Omega_{3}\sigma_{13}
-\imath\Omega_{4}\sigma_{15} -\frac{\Gamma_{4}+\gamma_{14} }{2}\sigma_{14}, \\
\dot{\sigma}_{15} &=& \imath(\delta_{1}-\delta_{2}-\delta_{3}+\delta_{4})\sigma_{15} +\imath\Omega^{\star}_{2}\sigma_{25} -\imath\Omega^{\star}_{4}\sigma_{14}
-\frac{\gamma_{15}}{2}\sigma_{15},\\
\dot{\sigma}_{23} &=& \imath\delta_{1}\sigma_{23} +\imath\Omega^{\star}_{2}\sigma_{13} -\imath\Omega^{\star}_{3}\sigma_{24}
+\imath\Omega^{\star}_{1}(\sigma_{33}-\sigma_{22}) -\frac{\Gamma_{2}+\gamma_{23}}{2}\sigma_{23}, \\
\dot{\sigma}_{24} &=& \imath(\delta_{1}-\delta_{3})\sigma_{24} -\imath\Omega_{3}\sigma_{23} -\imath\Omega_{4}\sigma_{25} +\imath\Omega^{\star}_{2}\sigma_{14}
+\imath\Omega^{\star}_{1}\sigma_{34} -\frac{\Gamma_{2}+\Gamma_{4}+\gamma_{24}}{2}\sigma_{24}, \\
\dot{\sigma}_{25} &=& \imath(\delta_{1}+\delta_{4}-\delta_{3})\sigma_{25} +\imath\Omega_{2}\sigma_{15} +\imath\Omega^{\star}_{1}\sigma_{35}
-\imath\Omega^{\star}_{4}\sigma_{24} -\frac{\Gamma_{2}+\gamma_{25}}{2}\sigma_{25},\\
\dot{\sigma}_{34} &=& -\imath\delta_{3}\sigma_{34} +\imath\Omega_{1}\sigma_{24} -\imath\Omega_{4}\sigma_{35} +\imath\Omega_{3}(\sigma_{44}-\sigma_{33})
-\frac{\Gamma_{4}+\gamma_{34}}{2}\sigma_{34},\\
\dot{\sigma}_{35} &=& -\imath(\delta_{3}-\delta_{4})\sigma_{35} -\imath\Omega_{3}\sigma_{45} +\imath\Omega_{1}\sigma_{25} -\imath\Omega^{\star}_{4}\sigma_{34}
-\frac{\gamma_{35}}{2}\sigma_{35}, \\
\dot{\sigma}_{45} &=& \imath\delta_{4}\sigma_{45} +\imath\Omega^{\star}_{4}(\sigma_{55}-\sigma_{44}) +\imath\Omega^{\star}_{3}\sigma_{35}
-\frac{\Gamma_{4}+\gamma_{45}}{2}\sigma_{45}.
\end{eqnarray}
\end{subequations}

\end{widetext}

\end{document}